\documentclass[11pt,a4paper]{iopart}
\usepackage[utf8]{inputenc}
\usepackage{url}
\usepackage{graphicx}
\usepackage{iopams}
\usepackage{setstack}
\usepackage{color}
\usepackage{bm}
\usepackage{accents}
\usepackage{hyperref}

\usepackage{cite}

\usepackage{wasysym}

\hypersetup{
    colorlinks=true,
    linkcolor=blue,
    filecolor=magenta,      
    citecolor=red
}

\usepackage{subfigure} 

\newcommand{\dd}{{\rm d}}
\newcommand{\udt}[3]{#1^{#2}_{\phantom{#2}#3}}
\newcommand{\udut}[4]{#1^{#2\phantom{#3}#4}_{\phantom{#2}#3\phantom{#4}}}

\newcommand{\dut}[3]{#1_{#2}^{\phantom{#2}#3}}
\newcommand{\dudt}[4]{#1_{#2\phantom{#3}#4}^{\phantom{#2}#3}}

\newcommand{\lc}[1]{\accentset{\circ}{#1}}

\makeatletter
\long\def\dddddot#1{%
  {\mathop {#1}\limits ^{\vbox to-1.4\ex@ {\kern -\tw@ \ex@ \hbox {\normalfont .....}\vss }}}%
}
\long\def\multidots#1#2{%
  \count@=0
  {{\mathop {#2}\limits ^{\vbox to-1.4\ex@ {\kern -\tw@ \ex@ \hbox {\normalfont %
  \loop%
  \ifnum#1>\count@%
  .%
  \advance\count@ by1%
  \repeat%
  }\vss }}}}%
}
\makeatother

\begin{document}

\title{\boldmath Well-Tempered Minkowski Solutions in Teleparallel Horndeski Theory}

\author{Reginald Christian Bernardo}
\address{National Institute of Physics, University of the Philippines Diliman, Quezon City 1101, Philippines.}
\ead{rbernardo@nip.upd.edu.ph}

\author{Jackson Levi Said}
\address{Institute of Space Sciences and Astronomy, University of Malta, Msida, MSD 2080, Malta.}
\address{Department of Physics, University of Malta, Msida, MSD 2080, Malta.}
\ead{jackson.said@um.edu.mt}

\author{Maria Caruana}
\address{Institute of Space Sciences and Astronomy, University of Malta, Msida, MSD 2080, Malta.}
\address{Department of Physics, University of Malta, Msida, MSD 2080, Malta.}
\ead{maria.caruana.16@um.edu.mt}

\author{Stephen Appleby}
\address{Asia Pacific Center for Theoretical Physics, Pohang, 37673, Korea.}
\address{Department of Physics, POSTECH, Pohang 37673, Korea.}
\ead{stephen.appleby@apctp.org}

\date{\today}

\begin{abstract}
\textbf{Well-tempering stands among the few classical methods of screening vacuum energy to deliver a late-time, low energy vacuum state. We build on the class of Horndeski models that admit a Minkowski vacuum state despite the presence of an arbitrarily large vacuum energy to obtain a much larger family of models in teleparallel Horndeski theory. We set up the routine for obtaining these models and present a variety of cases, all of which are able to screen a natural particle physics scale vacuum energy using degeneracy in the field equations. We establish that well-tempering is the unique method of utilizing degeneracy in Horndeski scalar-tensor gravity -- and its teleparallel generalisation -- that can accommodate self-tuned flat Minkowski solutions, when the explicit scalar field dependence in the action is minimal (a tadpole and a conformal coupling to the Ricci scalar). Finally, we study the dynamics of the well-tempered teleparallel Galileon. We generate its phase portraits and assess the attractor nature of the Minkowski vacuum under linear perturbations and through a phase transition of vacuum energy.}
\\ \\ \textit{``The effort to understand the Universe is one of the very few things that lifts human life a little above the level of farce, and gives it some of the grace of tragedy.'' - Steven Weinberg} 
\end{abstract}

\maketitle

\section{Introduction}
\label{sec:intro}

The standard model of cosmology is evidenced by an overwhelming amount of observational data describing the Universe at all scales where measurements are possible \cite{misner1973gravitation,Clifton:2011jh,Aghanim:2018eyx}. Recently, the effectiveness of this model in meeting these observational demands has been called into question with growing tensions in a number of cosmological parameters such as the value of the Hubble parameter at current times \cite{Bernal:2016gxb,DiValentino:2020zio,DiValentino:2021izs}. Within the standard model, the current phase of accelerated expansion \cite{Riess:1998cb,Perlmutter:1998np} is posited to be driven by the vacuum energy of spacetime. This is the source of the cosmological constant problem \cite{RevModPhys.61.1} where the vacuum energy density predicted by quantum field theory is in stark disagreement with the value measured through this acceleration within $\Lambda$CDM. This has prompted numerous possible solutions both in terms of additional modifications to the matter section beyond dark matter \cite{Baudis:2016qwx,Bertone:2004pz} and the cosmological constant \cite{Peebles:2002gy,Copeland:2006wr, Davidson:2009mp, Davidson:2014vda}. However, this has also led to a reconsideration of general relativity (GR) as the correct description of gravity in the standard model of cosmology, which has allowed a plethora of possible alternative scenarios \cite{CANTATA:2021ktz, Clifton:2011jh,Capozziello:2011et, Bamba:2012cp, Nojiri:2017ncd,Capozziello:2021bki,Capozziello:2020xem}.

The prospect of theories of gravity beyond GR that may be developed into observationally viable cosmological models has branched into several different directions. From the perspective of the Lovelock theorem \cite{Lovelock:1971yv}, these theories can be formed by adding arbitrary contributions of scalar invariants to the action \cite{DeFelice:2010aj,Capozziello:2011et}, adding extra fields to the theory (such as scalar fields) \cite{Kobayashi:2019hrl}, additional dimensions \cite{Dvali:2000hr}, considering non-local contributions \cite{Deser:2007jk}, or even by taking the prospect of an emergent form of gravity that does not come from an action formalism \cite{Witten:1998qj}. Throughout the decades, most promising theories have been shown to fall into the large class of models represented by Horndeski gravity \cite{Horndeski:1974wa}, which is the most general second-order theory of gravity containing one scalar field. Horndeski gravity offers the possibility of a rich phenomenology in which to construct cosmological models where the scalar field can take on a number of roles. 

This has led to strengthened efforts to resolve the cosmological constant problem in that the additional scalar field can be used to screen the spacetime deformation contribution of the cosmological constant. An unequivocal touchstone is a footnote in Steven Weinberg's seminal 1989 paper ``The cosmological constant problem'' \cite{RevModPhys.61.1} where the possibility of self-tuning fields was teased for the first time. In Ref.~\cite{Charmousis:2011bf}, it was realised that the Horndeski scalar field can be used to generate non-standard vacuum states, in which the metric relaxes to an exact Minkowski or de Sitter state, with the scalar field continuing to evolve on this background. Such behaviour requires a form of degeneracy in the field equations, which allows Poincar\'{e} invariance to be broken in the scalar sector. In this case, the scalar field can dynamically cancel the effect of an arbitrarily large vacuum energy, rendering the magnitude of the cosmological constant moot. Two different forms of degeneracy have been pursued in the literature; so-called Fab-Four \cite{Charmousis:2011bf} and well-tempering \cite{Appleby:2018yci}, and in this work we focus on the latter. Previous works have studied various phenomenological aspects of this model \cite{Emond:2018fvv,Bernardo:2021hrz}, with a particular emphasis on the regular Horndeski Fab Four class \cite{Copeland:2012qf, Starobinsky:2016kua, Torres:2018lni, Appleby:2015ysa,Appleby:2012rx}. Potentially viable cosmic histories can be constructed, with a standard matter epoch and late time de Sitter state that is shielded from the presence of an arbitrary vacuum energy. Some general conditions on the existence of degenerate vacuum states where provided in Ref.~\cite{Bernardo:2021hrz}.

The dynamical self tuning ideology of these works is not without its issues \cite{Linder:2013zoa,Starobinsky:2016kua}, and the generation of degenerate de Sitter states typically requires unpalatable derivative couplings between the scalar field and metric. In contrast, somewhat elegant Minkowski vacuum solutions can be obtained using the same approach, involving only the scalar field kinetic term, cubic Galileon and crucially the tadpole \cite{Appleby:2020dko}. Although modern cosmology is predicated on the existence of a low energy de Sitter state, Minkowski space provides a simple test bed that can be used to study the properties of degenerate vacuum solutions. In this work, we generalise the analysis of Ref.~\cite{Appleby:2020dko} to incorporate the enhanced model space afforded by the teleparallel gravity (TG) sector and its interactions with the scalar field. These contributions expand the model space in which degenerate Minkowski vacuum states exist, and also alter the dynamics of spacetime away from the vacuum. 

Horndeski gravity contains a large class of avenues in which to construct gravitational models. However, recent multimessenger observations have constrained the speed of gravitational waves to one part in $10^{15}$ which has severely limited the number of possible viable models in Horndeski gravity \cite{Ezquiaga:2017ekz}. We thus find ourselves with strong motivation to consider possible generalisations of Horndeski gravity. One of these possible revivals of Horndeski gravity is constructed by a change of geometry from curvature- to torsion-based theories through TG \cite{Bahamonde:2021gfp,Aldrovandi:2013wha,Cai:2015emx,Krssak:2018ywd,Capozziello:2018qcp}. Here, TG is the collection of all theories that embody torsion through the teleparallel connection \cite{Weitzenbock1923,Bahamonde:2021gfp} which is curvature-less and satisfies metricity. Analogous to the Ricci scalar $\lc{R}$ (over-circles denote all quantities calculated with the Levi-Civita connection), TG produces a torsion scalar $T$ which turns out to be equivalent to the regular Ricci scalar up to a boundary term. This is called the teleparallel equivalent to general relativity (TEGR), and is dynamically equivalent to GR.

TEGR can be generalised in a number of ways such as $f(T)$ gravity \cite{Ferraro:2006jd,Ferraro:2008ey,Bengochea:2008gz,Linder:2010py,Chen:2010va,Bahamonde:2019zea,Bamba:2012cp} which takes the same approach as $f(\lc{R})$ gravity \cite{Sotiriou:2008rp,Faraoni:2008mf,Capozziello:2011et}. In this work, our interest is in the various scalar-tensor \cite{Hohmann:2018vle,Hohmann:2018dqh,Hohmann:2018ijr, Chen:2021oal} forms of the theory since the naturally lower order nature of TG \cite{Gonzalez:2015sha,Bahamonde:2019shr} offers the possibility of a much larger family of analogous Horndeski models. Thus, a teleparallel analogue of Horndeski gravity (Teledeski) can be built \cite{Bahamonde:2019shr,Bahamonde:2020cfv,Bahamonde:2021dqn} which coincidentally revives previously disqualified Horndeski models due to some of the additional components of the theory appearing in the form of the speed formula of gravitational waves \cite{Bahamonde:2019ipm}.

In this work, we probe the space of well-tempered cosmologies that produce self-tuning models that yield a Minkowski vacuum. This complements previous work on de Sitter vacua \cite{Bernardo:2021izq}. We methodically analyze class of self-tuning cosmologies in Teledeski gravity to find explicit solutions that are well-tempered. We specifically focus on the impact of the Teledeski term on previous works in Minkowski vacua for regular Horndeski gravity \cite{Appleby:2020dko}. The  work proceeds as follows, we start with a brief introduction to Teledeski gravity and its field equations (Sec.~\ref{sec:teleparallel_horndeski}). Then, we present the well-tempered recipe (Sec.~\ref{sec:well_tempered_recipe}) through which we determine a variety of models admitting a well-tempered Minkowski vacuum, thereby expanding the range of the Fugue in B$\flat$ models that were previously obtained in Horndeski theory (Sec.~\ref{sec:well_tempered_minkowski}). We then show that well-tempering is the only means in the shift symmetric sector by which it is possible to exploit the degeneracy to obtain self-tuning flat Minkowski solutions (Sec.~\ref{sec:trivial_scalar_minkowski}). Lastly, we present the well-tempered teleparallel Galileon (Sec.~\ref{sec:wt_galileon}), studying various aspects of its dynamics on-shell and off-shell in a cosmological setting. We make our conclusions after this (Sec.~\ref{sec:conclusions}).

\textit{Conventions.} Geometrized units $c = M_{\rm{Pl}}^2 = 1$ will be used throughout where  $c$ is the speed of light in vacuum, $M_{\rm{Pl}}^2 = 1 / \left(8 \pi G\right)$ , and $G$ is Newton's gravitational constant. A dot over a variable means differentiation with respect to the cosmic time $t$ and a prime denotes the differentiation of a univariable function with respect to its argument. Over-circles refer to quantities based on the Levi-Civita connection. On tensors, Greek indices refer to coordinates on the general manifold while Latin ones refer to the local Minkowski spacetime. Derivatives of the Horndeski and Teledeski potentials with respect to the scalar field $\phi$ and the kinetic term $X$ are written as subscripts.

\section{A Teleparallel Horndeski Theory} \label{sec:teleparallel_horndeski}

The foundations on which Teledeski theory (Sec.~\ref{subsec:teledeski_theory}) is first presented which is then followed by the dynamical equations for background cosmology (Sec.~\ref{subsec:teledeski_cosmology}).

\subsection{A Teleparallel Gravity Toolkit}\label{subsec:teledeski_theory}

TG replaces curvature in GR by exchanging the curvature-full Levi-Civita connection $\udt{\lc{\Gamma}}{\sigma}{\mu\nu}$ (over-circles denote all quantities calculated with the Levi-Civita connection) with its analogue torsion-full teleparallel connection $\udt{\Gamma}{\sigma}{\mu\nu}$ \cite{Aldrovandi:2013wha,Bahamonde:2021gfp,Cai:2015emx,Krssak:2018ywd}. While the regular curvature-based tensors that are based on the Levi-Civita connection do not vanish ($\udt{\lc{R}}{\alpha}{\beta\gamma\epsilon}(\udt{\lc{\Gamma}}{\sigma}{\mu\nu}) \neq 0$), their teleparallel variants do ($\udt{R}{\alpha}{\beta\gamma\epsilon}(\udt{\Gamma}{\sigma}{\mu\nu}) \equiv 0$).

In practice, TG does this by building geometry on a gravitational tetrad $\udt{e}{A}{\mu}$ and a flat spin connection $\udt{\omega}{B}{C\nu}$. The tetrad then transforms inertial (Latin) indices to their general manifold (Greek) indices, as it does with the metric tensor
\begin{eqnarray}
    g_{\mu\nu} = \udt{e}{A}{\mu}\udt{e}{B}{\nu} \eta_{AB}\,, &\rm{and}& \eta_{AB} = \dut{E}{A}{\mu}\dut{E}{B}{\nu} g_{\mu\nu}\,,\label{eq:metr_trans}
\end{eqnarray}
which observes basic orthogonality conditions $\udt{e}{A}{\mu}\dut{E}{B}{\mu} = \delta_B^A$ and $\udt{e}{A}{\mu}\dut{E}{A}{\nu} = \delta_{\mu}^{\nu}$. On the other hand, the flat spin connection embodies the degrees of freedom associated with the local Lorentz transformation invariance. Together these two objects represents the degrees of freedom of the theory, which also together make up the teleparallel connection \cite{Weitzenbock1923,Cai:2015emx,Krssak:2018ywd}
\begin{equation}
    \udt{\Gamma}{\lambda}{\nu\mu}=\dut{E}{A}{\lambda}\partial_{\mu}\udt{e}{A}{\nu}+\dut{E}{A}{\lambda}\udt{\omega}{A}{B\mu}\udt{e}{B}{\nu}\,.
\end{equation}
The freedom in the Lorentz group means that there will be an infinite number of frames which satisfy the metric equation (\ref{eq:metr_trans}), but the one in which the spin connection is identically zero is called the Weitzenb\"{o}ck gauge \cite{Weitzenbock1923}.

Gravitational models are constructed in TG using contractions of the torsion tensor defined as \cite{Aldrovandi:2013wha,ortin2004gravity, Bahamonde:2021gfp}
\begin{equation}
\label{eq:torsion_tensor_def}
    \udt{T}{A}{\mu\nu} := 2\udt{\Gamma}{A}{[\nu\mu]}\,,
\end{equation}
which is analogous to curvature-based scenarios with the Riemann tensor, and where square brackets represent the antisymmetric operator. This tensor can be decomposed into irreducible parts \cite{PhysRevD.19.3524,Bahamonde:2017wwk}
\begin{eqnarray}
\label{eq:axial_torsion_irr}
    a_{\mu} &:=& \frac{1}{6}\epsilon_{\mu\nu\lambda\rho}T^{\nu\lambda\rho}\,,\\[4pt]
\label{eq:vector_torsion_irr}
    v_{\mu} &:=& \udt{T}{\lambda}{\lambda\mu}\,,\\[4pt]
\label{eq:tensor_torsion_irr}
    t_{\lambda\mu\nu} &:=& \frac{1}{2}\left(T_{\lambda\mu\nu}+T_{\mu\lambda\nu}\right)+\frac{1}{6}\left(g_{\nu\lambda}v_{\mu}+g_{\nu\mu}v_{\lambda}\right)-\frac{1}{3}g_{\lambda\mu}v_{\nu}\,,
\end{eqnarray}
which are the axial, vector, and purely tensorial parts, respectively, and where $\epsilon_{\mu\nu\lambda\rho}$ is the totally antisymmetric Levi-Civita tensor in four dimensions. This decomposition leads directly to the scalar invariants \cite{Bahamonde:2015zma}
\begin{eqnarray}
    T_{\rm{ax}} &:=& a_{\mu}a^{\mu} = -\frac{1}{18}\left(T_{\lambda\mu\nu}T^{\lambda\mu\nu}-2T_{\lambda\mu\nu}T^{\mu\lambda\nu}\right)\,,\\[4pt]
    T_{\rm{vec}} &:=& v_{\mu}v^{\mu}=\udt{T}{\lambda}{\lambda\mu}\dut{T}{\rho}{\rho\mu}\,,\\[4pt]
    T{_{\rm{ten}}} &:=& t_{\lambda\mu\nu}t^{\lambda\mu\nu}=\frac{1}{2}\left(T_{\lambda\mu\nu}T^{\lambda\mu\nu}+T_{\lambda\mu\nu}T^{\mu\lambda\nu}\right)-\frac{1}{2}\udt{T}{\lambda}{\lambda\mu}\dut{T}{\rho}{\rho\mu}\,,
\end{eqnarray}
which can be combined to produce the torsion scalar through \cite{Bahamonde:2021gfp}
\begin{eqnarray}
    T &:=& \frac{3}{2}T_{\rm{ax}}+\frac{2}{3}T_{\rm{ten}}-\frac{2}{3}T{_{\rm{vec}}}=\frac{1}{2}\left(E_{A}{}^{\lambda}g^{\rho\mu}E_{B}{}^{\nu}+2E_{B}{}^{\rho}g^{\lambda\mu}E_{A}{}^{\nu}+\frac{1}{2}\eta_{AB}g^{\mu\rho}g^{\nu\lambda}\right)T^{A}{}_{\mu\nu}T^{B}{}_{\rho\lambda}\,.\nonumber\\
\end{eqnarray}
This is an important scalar because it is equal to the Ricci scalar up to a boundary term, namely \cite{Bahamonde:2015zma}
\begin{equation}
    \lc{R}=-T+\frac{2}{e}\partial_{\mu}\left(e\udut{T}{\lambda}{\lambda}{\mu}\right):=-T+B\,,
\end{equation}
where $B$ is a total divergence term, and $e$ is the tetrad determinant. This guarantees that a linear torsion scalar will produce a TEGR due to the boundary term nature of $B$ \cite{Hehl:1994ue,Aldrovandi:2013wha}. This can be generalised to $f(T)$ gravity which is also second order, in an analogous fashion as $f(\lc{R})$ gravity is constructed \cite{Sotiriou:2008rp,Faraoni:2008mf,Capozziello:2011et}.

On the other hand, the scalar sector is dominated by the minimal coupling prescription in which partial derivatives are elevated to the Levi-Civita covariant derivative rather than a teleparallel covariant derivative \cite{Aldrovandi:2013wha,BeltranJimenez:2020sih} $\partial_{\mu} \rightarrow \mathring{\nabla}_{\mu}$, which applies to the whole matter sector. In this context, regular Horndeski gravity \cite{Horndeski:1974wa} can be elevated to a teleparallel setting by simply including some additional contributions to the Lagrangian \cite{Bahamonde:2019shr,Bahamonde:2019ipm,Bahamonde:2020cfv}.

Thus, we can write the regular Horndeski action along with an additional Teledeski term, namely 
\begin{equation}\label{action}
    \mathcal{S}_{\rm{BDLS}} = \frac{1}{2\kappa^2}\int d^4 x\, e\mathcal{L}_{\rm{Tele}} + \frac{1}{2\kappa^2}\sum_{i=2}^{5} \int d^4 x\, e\mathcal{L}_i+ \int d^4x \, e\mathcal{L}_{\rm m}\,,
\end{equation}
where the regular Horndeksi contributions appear as
\cite{Horndeski:1974wa}
\begin{eqnarray}
\mathcal{L}_{2} &:=& G_{2}(\phi,X)\,,\label{eq:LagrHorn1}\\[4pt]
\mathcal{L}_{3} &:=& -G_{3}(\phi,X)\mathring{\Box}\phi\,,\\[4pt]
\mathcal{L}_{4} &:=& G_{4}(\phi,X)\left(-T+B\right)+G_{4,X}(\phi,X)\left[\left(\mathring{\Box}\phi\right)^{2}-\phi_{;\mu\nu}\phi^{;\mu\nu}\right]\,,\\[4pt]
\mathcal{L}_{5} &:=& G_{5}(\phi,X)\mathring{G}_{\mu\nu}\phi^{;\mu\nu}-\frac{1}{6}G_{5,X}(\phi,X)\left[\left(\mathring{\Box}\phi\right)^{3}+2\dut{\phi}{;\mu}{\nu}\dut{\phi}{;\nu}{\alpha}\dut{\phi}{;\alpha}{\mu}-3\phi_{;\mu\nu}\phi^{;\mu\nu}\,\mathring{\Box}\phi\right]\,,\label{eq:LagrHorn5}
\end{eqnarray}
while the new Teledeski contribution gives
\begin{equation}
\label{eq:LTele}
    \mathcal{L}_{\rm{Tele}}:= G_{\rm{Tele}}\left(\phi,X,T,T_{\rm{ax}},T_{\rm{vec}},I_2,J_1,J_3,J_5,J_6,J_8,J_{10}\right)\,,
\end{equation}
where the kinetic term is defined as $X:=-\frac{1}{2}\partial^{\mu}\phi\partial_{\mu}\phi$, $\mathcal{L}_{\rm m}$ is the Jordan frame matter Lagrangian, $\kappa^2:=8\pi G$, $\lc{G}_{\mu\nu}$ is the standard Einstein tensor, and where semicolons represent Levi-Civita covariant derivatives.

For the scalar field $\phi$, the new scalar invariants are given by
\cite{Bahamonde:2019shr}
\begin{equation}
    I_2 = v^{\mu} \phi_{;\mu}\,,
\end{equation}
in terms of linear contractions with the torsion tensor, and 
\begin{eqnarray}
J_{1} &=& a^{\mu}a^{\nu}\phi_{;\mu}\phi_{;\nu}\,,\\[4pt]
J_{3} &=& v_{\sigma}t^{\sigma\mu\nu}\phi_{;\mu}\phi_{;\nu}\,,\\[4pt]
J_{5} &=& t^{\sigma\mu\nu}\dudt{t}{\sigma}{\alpha}{\nu}\phi_{;\mu}\phi_{;\alpha}\,,\\[4pt]
J_{6} &=& t^{\sigma\mu\nu}\dut{t}{\sigma}{\alpha\beta}\phi_{;\mu}\phi_{;\nu}\phi_{;\alpha}\phi_{;\beta}\,,\\[4pt]
J_{8} &=& t^{\sigma\mu\nu}\dut{t}{\sigma\mu}{\alpha}\phi_{;\nu}\phi_{;\alpha}\,,\\[4pt]
J_{10} &=& \udt{\epsilon}{\mu}{\nu\sigma\rho}a^{\nu}t^{\alpha\rho\sigma}\phi_{;\mu}\phi_{;\alpha}\,,
\end{eqnarray}
for quadratic contractions, and where semicolons denote Levi-Civita covariant derivatives. In the limit where $G_{\rm{Tele}} = 0$, we recover the regular Horndeski limit, as expected.

\subsection{Modified Friedmann Equations}
\label{subsec:teledeski_cosmology}

For the rest of this paper, we restrict our attention to the shift symmetric sector\footnote{
Shift symmetric in this context means $\phi$ dependence for which a constant shift $\phi \to \phi + c$ only modifies the existing mass scales in the model, $M_{\rm pl}^{2}$ and arbitrary vacuum energy $\rho_{\Lambda}$.} and write out the Horndeski potentials as
\begin{eqnarray}
\label{eq:V_ss_action} G_2\left(\phi, X \right) &=& V(X) - \lambda^3 \phi\,, \\
\label{eq:G_ss_action} G_3\left(\phi, X \right) &=& G(X)\,, \\
\label{eq:A_ss_action} G_4\left(\phi, X \right) &=& \frac{1}{2} \left( M_{\rm{Pl}}^2 + M \phi + A(X) \right) \,,
\end{eqnarray}
where $A(X)$ is an arbitrary function of the kinetic term $X$, and the Teledeski potential as
\begin{equation}
\label{eq:GTele_ss_action}
    \mathcal{L}_{\rm{Tele}} = \mathcal{G} \left( X, T, I_2 \right)\,,
\end{equation}
where $\lambda$ and $M$ are constants parametrizing the tadpole $\lambda^3 \phi$ and a conformal coupling $M \phi \lc{R}$ in the action. We leave the quintic Horndeski sector for future work.

We consider a spatially flat, homogeneous, and isotropic cosmology, or in terms of the metric, take the Friedmann–Lema\^{i}tre–Robertson–Walker (FLRW) line element  
\begin{equation}
    \dd s^2 = -N(t)^2 \dd t^2 + a(t)^2(\dd x^2 + \dd y^2 + \dd z^2)\,,
\end{equation}
for a scale factor $a(t)$ and lapse function $N(t)$, and which directly leads to the tetrad $\udt{e}{a}{\mu} = {\rm diag}(N(t),a(t),a(t),a(t))$ which is compatible with the Weitzenb\"{o}ck gauge \cite{Krssak:2018ywd,Bahamonde:2021gfp}.

By considering a perfect fluid with an energy density $\rho(t)$ and a pressure $\mathcal{P}(t)$ in the matter sector, taking appropriate variations, and assuming the lapse gauge ($N=1$), we obtain the Friedmann equation \cite{Bahamonde:2019shr}
\begin{equation}
    \mathcal{E}_{\rm Tele} + \sum_{i=2}^4 \mathcal{E}_i + \rho(t) = 0 \,,
\end{equation}
where

\begin{eqnarray}
    \mathcal{E}_{\rm Tele} &=& 6 H\dot{\phi}\mathcal{G}_{I_2}+12 H^2 \mathcal{G}_{T}+2X \mathcal{G}_{X}-\mathcal{G}\,,\\
    \mathcal{E}_2 &=& 2XV_{X} - V + \lambda^3 \phi \,,\\
    \mathcal{E}_3 &=& 6X\dot\phi HG_{X} \,,\\
    \mathcal{E}_4 &=& -3H^2 \left( M_{\rm{Pl}}^2 + M \phi + A( X ) \right) + 12H^2X(A_{X}+XA_{XX}) -3 M H\dot\phi \,,
\end{eqnarray}
where many of the Teledeski scalar invariants turn out to vanish at background level. In this setting, the Hubble parameter is given as $H = \dot{a}/a$, while the kinetic term is denoted by $X = \dot{\phi}^2 / 2$, while the TG scalars are given by $T = 6H^2$ and $I_2 = 3H\dot{\phi}$. The second Friedmann equation is then given by
\begin{equation}
    \mathcal{P}_{\rm Tele}+\sum_{i=2}^4 \mathcal{P}_i + \mathcal{P}(t) = 0 \,,
\end{equation}
where
\begin{eqnarray}
    \mathcal{P}_{\rm Tele} &=&-3 H\dot{\phi} \mathcal{G}_{I_2}-12 H^2 \mathcal{G}_{T}-\frac{d}{dt}\Big(4H \mathcal{G}_{T}+\dot{\phi}\,\mathcal{G}_{I_2}\Big)+\mathcal{G}\,,\\
    \mathcal{P}_2 &=& \, V - \lambda^3 \phi  \,,\\
    \mathcal{P}_3 &=& -2X \ddot \phi G_{X} \,,\\
    \mathcal{P}_4 &=& \left(3H^2+2\dot H\right) \left( M_{\rm{Pl}}^2 + M \phi + A - 2 X A_X \right) \nonumber \\
& & - 2 H\dot X A_{X} - 4 H X \dot X A_{XX} + M \left(\ddot\phi+2H\dot\phi\right) \, .
\end{eqnarray}
In tandem, the scalar field produces a generalised Klein-Gordon equation of the type
\begin{equation}
    \frac{1}{a^3}\frac{d}{dt}\Big[a^3 (J+J_{\rm Tele})\Big]=P_{\phi}+P_{\rm Tele}\,,
\end{equation}
where the standard Horndeski terms appear as $J$ and $P_{\phi}$, given by \cite{Kobayashi:2011nu}
\begin{eqnarray}
    J &=& \dot\phi V_{X}+6HXG_{X} + 6H^2\dot\phi\left(A_{X}+2XA_{XX}\right) \,,\\
    P_{\phi} &=& -\lambda^3 + 3M\left(2H^2+\dot H\right) \,,
\end{eqnarray}
while $J_{\rm Tele}$ and $P_{\rm Tele}$ are additional Teledeski terms represented by
\begin{eqnarray}
    J_{\rm Tele} &=& \dot{\phi} \mathcal{G}_{X}\,,\\
    P_{\rm Tele} &=& -9 H^2\mathcal{G}_{I_2}+\mathcal{G}_{\phi}-3  \frac{d}{dt}\left(H\mathcal{G}_{I_2}\right)\,.
\end{eqnarray}
We shall use these equations for the remainder of the paper to determine the sectors of the theory accommodating a Minkowski vacuum with a self-tuning scalar field $\phi(t)$.

\section{The Well-Tempered Minkowski Recipe}
\label{sec:well_tempered_recipe}

We describe well-tempering and setup the necessary ingredients to pin down the sectors of Teledeski gravity that accommodates this desirable feature.

Well-tempering is a dynamical screening mechanism designed to protect the spacetime from the influence of an arbitrarily large vacuum energy \cite{Appleby:2018yci, Appleby:2020njl}. This self-tuning mechanism meets two particular advantages: it does not rely on the tuning of mass scales in the action, and is shown to be compatible with the existence of matter-dominated universes. The dynamical cancellation of vacuum energy happens in the vacuum state in the degenerate space of the dynamical equations.

Minkowski well-tempered solutions can be obtained as follows. The Hubble and scalar field equation can be generically written in the forms
\begin{equation}
\label{eq:Heq_generic}
\dot{H} = \ddot{\phi} \ \mathcal{Z} \left( \phi, \dot{\phi}, H \right) + \mathcal{Y}\left( \phi, \dot{\phi}, H \right)\,,
\end{equation}
and
\begin{equation}
\label{eq:Seq_generic}
0 = \ddot{\phi} \ \mathcal{D} \left( \phi, \dot{\phi}, H \right) + \mathcal{C} \left( \phi, \dot{\phi}, H , \dot{H} \right) \,,
\end{equation}
where the functions $\mathcal{Y}$, $\mathcal{Z}$, $\mathcal{C}$, and $\mathcal{D}$ are determined by the theory in consideration. See Eqs.~(\ref{eq:mathcalY}), (\ref{eq:mathcalZ}), (\ref{eq:mathcalC}), (\ref{eq:mathcalD}) for their Teledeski functional forms in a Minkowski vacuum. First, evaluate the dynamical equations on a Minkowski vacuum (or \textit{on-shell}), i.e., $P_\Lambda = -\rho_\Lambda$ and $H = 0$, $\dot{H} = 0$, where the matter pressure and energy density are now given by constants $P(t) = P_\Lambda$ and $\rho(t) = \rho_\Lambda$, respectively.

This step over-constrains the dynamical system, because we have arbitrarily fixed the dynamics of $H$ with an ansatz. Well-tempering resolves this issue by utilizing the degeneracy in the field space of the dynamical equations. In particular, Eqs.~(\ref{eq:Heq_generic}) and (\ref{eq:Seq_generic}) can be made to be equivalent on a Minkowski vacuum provided that
\begin{equation}
\label{eq:degeneracy_equation}
\mathcal{Y} \mathcal{D} - \mathcal{C} \mathcal{Z} = 0 , \ \ \ \ \rm{on-shell}  ,
\end{equation}
when $H=0$, $\dot{H} = 0$. We refer to Eq.~(\ref{eq:degeneracy_equation}) as the degeneracy equation. Then, if the degeneracy equation is satisfied and the scalar field explicitly appears in the Hamiltonian constraint (Friedmann equation) on-shell
\begin{equation}
\label{eq:consistency_3}
\rho_\Lambda =  F \left[ \phi(t), \dot{\phi}\left(t\right) \right] , \ \ \ \ \rm{on-shell} ,
\end{equation}
for some function $F$, we are guaranteed a self-tuning Minkowski vacuum in the resulting theory. This recipe leads to a \textit{fugue} of Horndeski functions accommodating a well-tempered Minkowski solution. We now expand this to Teledeski theory, using the supplementary freedom that TG bestows in scalar-tensor theory.

The on-shell coefficients $\mathcal{Y}$, $\mathcal{Z}$, $\mathcal{C}$, and $\mathcal{D}$ in the field equations can be identified as
\begin{eqnarray}
\label{eq:mathcalY} \mathcal{Y} &=& 3 V_X \dot{ \phi }^2 + 3 \mathcal{G}_X \dot{\phi}^2\,,  \\
\label{eq:mathcalZ} \mathcal{Z} &=& 3 M - 3 G_X \dot{ \phi }^2 -3 \mathcal{G}_{I_2} - 3 \mathcal{G}_{X I_2} \dot{ \phi }^2\,,  \\
\label{eq:mathcalC} \mathcal{C} &=& - \lambda^3\,, \\
\label{eq:mathcalD} \mathcal{D} &=& - V_X -V_{XX} \dot{ \phi }^2 - \mathcal{G}_X - \mathcal{G}_{XX} \dot{ \phi }^2 \,.
\end{eqnarray}
We note that $T = I_2 = 0$ on-shell, as expected because the torsion is gravitational in nature and so it vanishes as well as the other gravitational scalars that can be obtained with it. Furthermore, all of the terms arising from the revived Horndeski potential $A(X)$ vanish in the Minkowski limit, because this Horndeski sector is of purely gravitational nature. Substituting the above coefficients into Eq.~(\ref{eq:degeneracy_equation}), we obtain
\begin{equation}
\label{eq:BFlatM7}
\lambda ^3 M=2 X \left(\left(\mathcal{G}_X+V_X\right) \left(2 X \left(\mathcal{G}_{XX}+V_XX\right)+\mathcal{G}_X+V_X\right)+\lambda ^3 \mathcal{G}_{X I_2}\right)+\lambda ^3 \mathcal{G}_{I_2}+2 \lambda ^3 X G_X \, .
\end{equation}
This degeneracy equation is the generalisation to Eqs.~(3.3) [\textit{B Flat Minor}] and (4.2) [\textit{B Flat Major}] in Ref.~\cite{Appleby:2020dko}. Now, it is given in terms of the Horndeski potentials ($V$ and $G$) and the Teledeski potential ($\mathcal{G}$). The interplay between $V$, $G$, and $\mathcal{G}$ allows the expansion of the Fugue of scalar field potentials that permit the existence of a well-tempered Minkowski vacuum.

We proceed in the next section to explore various solutions of the degeneracy equation.

\section{Well-Tempered Minkowski Solutions: Teleparallel Extensions}
\label{sec:well_tempered_minkowski}

We briefly review the established Horndeski solutions to Eq. (\ref{eq:BFlatM7}) in Sec. \ref{subsec:BFlat} and present new solutions in TG in Sec. \ref{subsec:BFlatM7}.

\subsection{Fugue in B Flat: The Original Horndeski Cover}
\label{subsec:BFlat}

In \textit{B Flat minor} (B$\flat$m: $M = 0$ and $\mathcal{G} = 0$), the degeneracy equation becomes
\begin{equation}
G_X = -\frac{ V_X \left(V_X + 2 X V_{XX}\right)}{\lambda ^3} \,.
\end{equation}
Since this is presented in terms of two potentials, this can be solved by specifying either one and then solving the differential equation for the other. To illustrate the methodology, a canonical kinetic term in $V$ can be substituted. This leads to the pair of Horndeski potentials
\begin{eqnarray}
V(X) &=& \epsilon X\,, \\
G(X) &=& - \frac{ \epsilon^2 X }{\lambda^3}\,,
\end{eqnarray}
where $\epsilon \neq 0$ is a dimensionless constant. In this theory, the scalar field and the Hamiltonian constraint on-shell can be shown to be $\phi(t) = c_1 + c_2 t - t^2 \lambda^3 / \left(2\epsilon\right)$ and $\rho_\Lambda + \lambda^3 c_1 +\epsilon c_2^2 /2 = 0$. This reveals that vacuum energy $\rho_\Lambda$ is being screened by the pair of integration constants $\left(c_1, c_2\right)$ which characterizes the dynamics of the scalar field. The metric is exactly Minkowski space by design. General power law potentials can also be admitted
\begin{eqnarray}
V(X) &=& \frac{\epsilon  X^{n+1}}{n+1}\,, \\
G(X) &=& -\frac{\epsilon ^2 X^{2 n+1}}{\lambda ^3} \,,
\end{eqnarray}
where $\epsilon$ is now a dimension-full constant. This time, the on-shell scalar field can be written as $\dot{\phi} = \left( - \left( 2^n \lambda^3 t/ \epsilon \right) + c_1 \right)^{1/(2n + 1)}$ and the Hamiltonian constraint becomes $\rho_\Lambda + \lambda^3 c_2 = 0$ where $\phi(t) \sim c_2$.

In \textit{B Flat Major} (B$\flat$M: $\mathcal{G} = 0$), the degeneracy equation becomes
\begin{equation} \label{eq:deg_eq_horndeski}
\frac{2 X V_X}{M-2 X G_X}=\frac{\lambda ^3}{V_X + 2 X V_{XX}}\,.
\end{equation}
A canonical kinetic term ansatz to this leads to
\begin{eqnarray}
V(X) &=& \epsilon X \,,\\
G(X) &=& - \frac{ \epsilon^2 X }{\lambda^3} + \frac{M}{2} \ln(X/X_0)\,,
\end{eqnarray}
where $X_0$ is a constant. This can also be easily generalised to 
\begin{eqnarray}
\label{eq:V_BFlatM} V(X) &=& \frac{\epsilon  X^{n+1}}{n+1}\,, \\
\label{eq:G_BFlatM} G(X) &=& -\frac{\epsilon ^2 X^{2 n+1}}{\lambda ^3} + \frac{M}{2} \ln(X/X_0) \,.
\end{eqnarray}

Clearly, these B$\flat$m and B$\flat$M potentials differ in the added braiding $(M/2) \ln(X/X_0)$ which is the feedback of well-tempering to the added conformal coupling to the Ricci scalar. It is also worth noting that the scalar field and the Hamiltonian constraint on-shell are the same in both B$\flat$m and B$\flat$M due to the scalar field equation being fully characterized by $V$ -- the degeneracy equation fixes $G$ in terms of $V$.

\subsection{Fugue in B Flat: The Teleparallel Gravity Cover}
\label{subsec:BFlatM7}

In this section, we present new B$\flat$ models opened up by contributions from TG. All of these models are solutions to the degeneracy equation (\ref{eq:degeneracy_equation}) and hence can screen an arbitrarily large cosmological constant and possess an exact Minkowski vacuum.

To start, we remind ourselves that on-shell, any gravitational couplings vanish and $T = I_2 = 0$. Two important implications can be realised from this and the form of the degeneracy equation (\ref{eq:degeneracy_equation}). First, there should be no non-analytic dependence on $T$ and $I_2$; otherwise, the field equations will become undefined. Second, only terms linear in $I_2$, i.e., $\mathcal{G} \sim I_2$, can possibly influence the behavior of the scalar field on-shell. This can be realised since the highest derivative of the Teledeski potential $\mathcal{G}$ in $I_2$ in the degeneracy equation (\ref{eq:degeneracy_equation}) is first-order. In addition to a linear $I_2$ term, any higher-order analytic function in $I_2$ that vanish in the Minkowski limit $T, I_2 \rightarrow 0$ can be present. Based on these considerations, we can write down the most general Teledeski potential that leaves a nontrivial contribution to Eq.~(\ref{eq:BFlatM7}) as
\begin{equation}
\label{eq:FugueBFlatM7}
    \mathcal{G}(X, T, I_2) = \tilde{ \mathcal{G} }(X) \left( I_2 + F(T, I_2) \right)\,,
\end{equation}
where $\tilde{ \mathcal{G}}$ on the right hand side is an arbitrary function of $X$ and $F(T, I_2) = \mathcal{O}( T, I_2^2 )$ in the limit $T , I_2 \rightarrow 0$ but is otherwise an arbitrary analytic function. A simple way to understand this is to take a term $F(T, I_2) \sim T I_2^2$. By substituting Eq.~(\ref{eq:FugueBFlatM7}) into the degeneracy equation (\ref{eq:BFlatM7}), one finds that all contributions from $F$ vanishes precisely because they vanish in the limit $T , I_2 \rightarrow 0$.

We emphasize Eq.~(\ref{eq:FugueBFlatM7}) as an important result. The \textit{nearly-free} function $F$ embodies a huge boost in the theory space through which well-tempered models can be adjusted \textit{off-shell} for phenomenological purposes. This will play a major role later in our analysis of the dynamics of the well-tempered teleparallel Galileon in Sec.~\ref{sec:wt_galileon}. The discussion also brings us back to the $\mathcal{L}_4(X)$ (revived Horndeski) terms which vanished in the Minkowski limit because they enter the field equations besides factors of the Hubble function. Therefore, an arbitrary function $A(X)$ representative of the revived Horndeski sector can also be accepted in the well-tempered model. It must be noted that $A$ does not even necessarily have to be analytic since it does not depend on an argument that vanishes in the Minkowski limit.

Moving on, we substitute Eq.~(\ref{eq:FugueBFlatM7}) into Eq.~(\ref{eq:BFlatM7}) to obtain
\begin{equation}
\label{eq:BFlatM7_recast}
6 X V_X \left(2 X V_{XX} + V_X \right)=3 \lambda ^3 \left(M -2 X G_X - 2 X \tilde{\mathcal{G}}_X - \tilde{\mathcal{G}} \right) \,.
\end{equation}
This degeneracy equation is described by the potentials $V$, $G$, and $\tilde{\mathcal{G}}$. The interplay between these three potentials allows for the wider range of possibilities extending the B Flat models to the Teledeski sector. We refer collectively to the new models as \textit{Fugue in B Flat: The Teleparallel Gravity Cover} (B$\flat$TG).

It can be shown that given Eq.~(\ref{eq:BFlatM7_recast}), the scalar equation and the Hamiltonian constraint on-shell can be written as
\begin{equation}
\label{eq:SeqBFlatM7}
\ddot{\phi} \left( V_X + \dot{\phi}^2 V_{XX} \right) + \lambda^3 = 0\,,
\end{equation}
and
\begin{equation}
\label{eq:FeqBFlatM7}
\rho_\Lambda = V - \dot{\phi}^2 V_X -  \lambda^3 \phi(t)\,.
\end{equation}
The Hamiltonian constraint confirms that in B$\flat$TG the vacuum energy is being dynamically cancelled. Notably, the above expressions turn out to depend only on $V$ as a consequence of well-tempering. However, in B$\flat$TG, specifying $V$ no longer completely determines the braiding $G$, but more generally, because of the expanded freedom allowed in TG, leads to a $(G, \tilde{\mathcal{G}})$-family of functions, all of which permit the existence of a well-tempered Minkowski vacuum in the model. Thus, in practice, any two of the potentials among $V$, $G$, and $\tilde{\mathcal{G}}$ can be chosen to obtain a B$\flat$TG model. We proceed with a familiar model to demonstrate this procedure.

Suppose we consider the Galileon-type Horndeski potentials
\begin{eqnarray}
\label{eq:V_gal} V(X) &=& \alpha X\,,\\
\label{eq:G_gal} G(X) &=& \gamma X\,,
\end{eqnarray}
where $\alpha$ and $\gamma$ are constants. By substituting this into Eq.~(\ref{eq:BFlatM7_recast}) and then solving the resulting differential equation, we obtain the Teledeski potential
\begin{equation}
\label{eq:mathcalG_gal}
    \tilde{\mathcal{G}}(X) = \frac{g}{\sqrt{X}}+M -\frac{2 \left(\alpha ^2+\gamma  \lambda ^3\right)}{3 \lambda ^3} X\,,
\end{equation}
where $g$ is an arbitrary, free constant that can be set to zero. We refer to this as the \textit{well-tempered teleparallel Galileon}. This shows the possibility of obtaining simple and elegant well-tempered models where the constant $\alpha$ and mass scale $\gamma$ can take order unity and natural particle physics energy scales respectively. In other words, no fine-tuning of mass scales is required. This motivates the consideration of a well-tempered Minkowski solution to screen vacuum energy. Furthermore, the off-shell dynamics being controlled by an arbitrary Teledeski potential (Eq.~(\ref{eq:FugueBFlatM7})) is clearly an asset for phenomenology. Later, we shall study the dynamics of this model in Sec.~\ref{sec:wt_galileon}.

We can also further obtain a condensate-type model:
\begin{eqnarray}
\label{eq:V_condensate} V(X) &=& \alpha X + \beta X^2\,, \\
\label{eq:G_condensate} G(X) &=& \gamma X\,,
\end{eqnarray}
where $\alpha$, $\beta$, and $\gamma$ are constants. This is often referred to as the Galileon ghost condensate \cite{Kase:2018aps} and can be regarded as a competetive theory as far as the existing cosmological data is concerned \cite{Peirone:2019aua, Frusciante:2020zfs}. By substituting this into Eq.~(\ref{eq:BFlatM7_recast}), we can solve for the Teledeski potential
\begin{equation}
\label{eq:mathcalG_condensate}
\tilde{\mathcal{G}}(X) = \frac{g}{\sqrt{X}}+M -\frac{2 \left(\alpha ^2+\gamma  \lambda ^3\right)}{3 \lambda ^3} X - \frac{16 \alpha  \beta }{5 \lambda ^3} X^2 - \frac{24 \beta ^2 }{7 \lambda ^3} X^3 \,.
\end{equation}
We refer to this (Eqs.~(\ref{eq:V_condensate}), (\ref{eq:G_condensate}), and (\ref{eq:mathcalG_condensate})) as the \textit{well-tempered teleparallel condensate}. In this model, it can be shown that the on-shell scalar field is given by
\begin{equation}
\dot{\phi} = \frac{2^{1/3} \omega(t)^{2/3}-\left( 2 \times 3^{1/3} \right) \alpha  \beta }{6^{2/3} \beta \omega(t)^{1/3} } \,,
\end{equation}
where
\begin{equation}
\omega(t) = \sqrt{3 \beta ^3 \left(4 \alpha ^3+27 \beta  \left(c_1-\lambda ^3 t\right){}^2\right)}+9 \beta ^2 \left(c_1-\lambda ^3 t\right)\,,
\end{equation}
where $c_1$ is a constant. The Hamiltonian constraint on-shell can be expressed as
\begin{equation}
\rho_\Lambda +\lambda ^3 \phi (t) + \alpha X + 3 \beta X^2 = 0\,,
\end{equation}
showing the dynamical cancellation of vacuum energy that is taking place in the well-tempered Minkowski vacuum.\medskip

To end the section, we point out that many exact models can be sought by solving Eq.~(\ref{eq:BFlatM7_recast}). We present well-tempered models generalising the aesthetic Galileon class in the \ref{sec:non_galileon_models}. All of these perform the same fundamental task; they well-temper and protect the spacetime from responding to an arbitrarily large vacuum energy. Obviously, some of these, in particular, the ones involving the exponential and $\arctan$ couplings, are a lot more computationally expensive to study. This realisation adds practical value to the Galileon-type well-tempered models. We carry this mindset to Sec.~\ref{sec:wt_galileon} where we study the dynamics of the well-tempered teleparallel Galileon.

\section{Trivial Scalar and Other Minkowski Solutions}
\label{sec:trivial_scalar_minkowski}

In this section we show that there are no other kinds of self-tuning for a flat Minkowski spacetime metric except well-tempering, if there is no explicit $\phi$ dependence in the action beyond the tadpole $\lambda^3 \phi$ and the conformal coupling $M \phi \lc{R}$ [Eqs.~(\ref{eq:V_ss_action}-\ref{eq:GTele_ss_action})]. We do so by exploring the alternative ways of exploiting degeneracy in the field equations or rather possible caveats to the recipe given in Sec.~\ref{sec:well_tempered_recipe}.

Degeneracy requires that either one of Eqs.~(\ref{eq:Heq_generic}, \ref{eq:Seq_generic}) is trivially zero when we impose the metric ansatz $H=0$, $\dot{H} = 0$, or they are equivalent. We first consider the \textit{trivial scalar} approach in which the scalar field equation becomes identically satisfied on-shell; in symbols, $\mathcal{D} = \mathcal{C} = 0$. This leads to the progenitor self-tuning models widely-known as the \textit{Fab Four} in Horndeski gravity. In the shift symmetric Teledeski theory (Eq.~(\ref{action})), we find that $\mathcal{C} = 0$ implies that $\lambda = 0$ and hence there must be no tadpole. On the other hand, $\mathcal{D} = 0$ can be written as
\begin{equation}
\left( V_X +  \mathcal{G}_X \right) + \dot{ \phi }^2 \left( V_{XX} + \mathcal{G}_{XX}  \right) = 0\,.
\end{equation}
By recalling that $T = I_2 = 0$, we recognize that this equation can be satisfied by any analytic function of $\mathcal{G}$ that is at least linear in the torsion scalar $T$ and the scalar-torsion coupling $I_2$. The most general nontrivial Teledeski potential that can be considered is therefore of the form
\begin{equation}
\label{eq:mathcalG_trivial_scalar}
\mathcal{G}(X, T, I_2) = Y(X) \left( 1 + U(T, I_2) \right) \,,
\end{equation}
where $U(T, I_2) = \mathcal{O}(T, I_2)$ but is otherwise completely arbitrary. The leading term $\mathcal{G}(X, T, I_2) \sim Y(X)$ can be absorbed into $V(X)$. The $\mathcal{G}$ contribution to $\mathcal{D} = 0$ therefore vanishes and inevitably leads to the Horndeski potential $V$ given by
\begin{equation}
\label{eq:cuscuton}
V(X) = b_1 \sqrt{X} + b_2\,,
\end{equation}
where $b_1$ and $b_2$ may be identified as mass scales in the action. However, the Hamiltonian constraint on-shell becomes $\rho_\Lambda = b_2$ showing that vacuum energy is being cancelled by an additional cosmological constant component $b_2$ rather than by any dynamical effect.

We move on to a second possible caveat to the recipe: $\mathcal{Y} = \mathcal{C} = 0$. In this case, both Eqs.~(\ref{eq:Heq_generic}, \ref{eq:Seq_generic}) would reduce to $\ddot{\phi} = 0$. As in the trivial scalar approach, $\mathcal{C} = 0$ implies that $\lambda = 0$ and no tadpole. On the other hand, $\mathcal{Y}  = 0$ can be written as
\begin{equation}
V_X + \mathcal{G}_X = 0 \,.
\end{equation}
As above, this suggests that the most general nontrivial Teledeski potential is also given by Eq. (\ref{eq:mathcalG_trivial_scalar}) and $\mathcal{G}$ can be absorbed into $V$. We can then solve $V(X) = b_2$ and the Hamiltonian constraint on-shell reads $\rho_\Lambda = b_2$, showing that no dynamical cancellation of vacuum energy occurs.

As another possible caveat to the recipe, we consider $\mathcal{Y} = \mathcal{D} = 0$. We can write down this degeneracy condition as
\begin{equation}
\left( V_X +  \mathcal{G}_X \right) + \dot{ \phi }^2 \left( V_{XX} + \mathcal{G}_{XX}  \right) = 0\,,
\end{equation}
\begin{equation}
V_X + \mathcal{G}_X = 0 \,.
\end{equation}
Again, this suggests the general form of Eq.~(\ref{eq:mathcalG_trivial_scalar}) for the Teledeski potential, leads to $V(X) = b_2$, and the Hamiltonian constraint on-shell $\rho_\Lambda = b_2$. It must be noted that $\mathcal{D} = 0$ also automatically implies that $\mathcal{C} = \lambda^3 = 0$ or else the system cannot be consistent. Therefore, no dynamical cancellation of the vacuum energy occurs.

Lastly, we consider a \textit{trivial Hubble} approach or, in symbols, $\mathcal{Y} = \mathcal{Z} = 0$ where the Hubble equation becomes trivially satisfied on-shell. The equations that must be satisfied are
\begin{equation}
V_X + \mathcal{G}_X = 0 \,,
\end{equation}
\begin{equation}
3 M - 3 G_X \dot{ \phi }^2 -3 \mathcal{G}_{I_2} - 3 \mathcal{G}_{X I_2} \dot{ \phi }^2 = 0 \,.
\end{equation}
Due to $T = I_2 = 0$ and the appearance of the $I_2$ derivatives in $\mathcal{Z}$, we find that the general nontrivial Teledeski potential that can be accommodated is given by Eq.~(\ref{eq:FugueBFlatM7}). This leads to $V(X) = b_2$ and also inevitably implies that $\mathcal{D} = 0$. In fact, $\mathcal{Y} = 0$ always leads to $\mathcal{D} = 0$ which consequently implies $\mathcal{C} = 0$ or that there must be no tadpole term in the Lagrangian. Therefore, the Hamiltonian constraint becomes $\rho_\Lambda = b_2$ on-shell showing no dynamical cancellation of the vacuum energy is at work in this scenario.

This section therefore supports the conclusion: If there is no explicit $\phi$ dependence in the action beyond the tadpole $\lambda^3 \phi$ and the conformal coupling $M \phi \lc{R}$ [Eqs.~(\ref{eq:V_ss_action}-\ref{eq:GTele_ss_action})], then well-tempering is the only way forward utilizing the degeneracy in field space to obtain a self-tuning Minkowski vacuum.

We stress that this conclusion does not contradict the Fab Four which was constructed to generally admit Milne spacetimes \cite{Charmousis:2011bf, Copeland:2012qf}. As a matter of fact, we suspect self-tuning to occur in one of the caveats discussed in this section in a Milne slicing. We intend to discuss this in detail in a different work.

\section{The Well-Tempered Teleparallel Galileon}
\label{sec:wt_galileon}

We now study the cosmological dynamics of a model (Sec. \ref{subsec:theory_wt_galileon}) endowed with a well-tempered Minkowski vacuum. We particularly look at its linear dynamical stability (Sec. \ref{subsec:linear_stability}), phase portraits (Sec. \ref{subsec:phase_portraits}), and response to a phase transition of vacuum energy (Sec. \ref{subsec:phase_transition}).

\subsection{Theory}
\label{subsec:theory_wt_galileon}

We consider the Horndeski potentials
\begin{eqnarray}
\label{eq:V_galileon} V(X) &=& \alpha X\,, \\
\label{eq:G_galileon} G(X) &=& \gamma X\,, \\
\label{eq:A_galileon} A(X) &=& \zeta X^2\,,
\end{eqnarray}
and the Teledeski potential
\begin{eqnarray}
\label{eq:mathcalG_galileon}
\mathcal{G}(X, T, I_2) = \left( \frac{g}{\sqrt{X}}+M -\frac{2 \left(\alpha ^2+\gamma  \lambda ^3\right)}{3 \lambda ^3} X \right) \left( I_2 + \sigma T I_2^2 \right)\,,
\end{eqnarray}
where $\alpha$, $\gamma$, $\zeta$, $g$, and $\sigma$ are constants. The gravitational action for the theory is given by
\begin{eqnarray}
S_g \left[ e_{\ \mu}^A, \phi \right] &=& \int d^4x e 
\bigg[
 \alpha X -\lambda^3 \phi - \gamma X \mathring{\Box} \phi +\left( \frac{ M_{\rm{Pl}}^2 + M \phi + \zeta X^2 }{ 2 } \right) (-T + B) \nonumber\\
& & + 2 \zeta X\left[\left(\mathring{\Box}\phi\right)^{2}-\phi_{;\mu\nu}\phi^{;\mu\nu}\right] \nonumber\\
& & + \left( \frac{g}{\sqrt{X}}+M -\frac{2 \left(\alpha ^2+\gamma  \lambda ^3\right)}{3 \lambda ^3} X \right) \left( T^{\lambda \ \mu}_{\ \lambda} \phi_{; \mu} + \sigma T \left( T^{\lambda \ \mu}_{\ \lambda} \phi_{; \mu} \right)^2 \right) \bigg] \,.\nonumber\\
\label{eq:action_wt_galileon}
\end{eqnarray}
We refer to this as the well-tempered teleparallel Galileon due to its Horndeski sector being of the form of the widely-known Galileon while meeting the demands of well-tempering through the inclusion of TG terms.

The Friedmann equation can be shown to be
\begin{eqnarray}
3 H^2 \left( 1 + M \phi \right) &=&
\rho  + \lambda^3 \phi +\frac{\alpha  \dot{\phi}^2}{2}  -\left( \gamma + \frac{ \alpha ^2  }{\lambda^3} \right) \left(126 \sigma  H^4 \dot{\phi}^4 + 3 H \dot{\phi}^3 \right) \nonumber\\
& & + 3 \gamma H \dot{\phi}^3 + 54 \sigma  H^4 \dot{ \phi } \left(4 \sqrt{2} g+5 M \dot{\phi} \right) +\frac{45}{4} \zeta  H^2 \dot{\phi}^4 \,.\label{eq:Feq_galileon}
\end{eqnarray}
On the other hand, the Hubble and scalar field equations are given by
\begin{eqnarray}
2 \left( 1 + M \phi \right) \dot{H}
&=& -\rho  - P - 2 M H \dot{\phi} -\alpha \dot{ \phi }^2  - \gamma \dot{\phi}^2 \left( 3 H \dot{\phi} - \ddot{\phi} \right) \nonumber\\
& & +\frac{3}{2} \zeta \dot{ \phi }^3 \left( \dot{H} \dot{\phi} + 4 H \ddot{\phi} - 6 H^2 \dot{\phi} \right) \nonumber\\
& & -18 \sigma  H^2 \left(-12 \dot{H} \dot{\phi} \left(\sqrt{2} g+M \dot{\phi} \right)+ \left( 3 H^2 \dot{\phi} - 4 H \ddot{\phi} \right) \left(\sqrt{2} g+2 M \dot{\phi} \right) \right) \nonumber\\
& & +\left( \gamma + \frac{\alpha^2}{\lambda^3} \right) \left(24 \sigma  H^2 \dot{\phi}^3 \left(3 \left(H^2-\dot{H}\right) \dot{\phi} - 4 H \ddot{\phi} \right)+\dot{\phi}^2 \left(3 H \dot{\phi}-\ddot{\phi}\right)\right) \,,\nonumber\\
&&\label{eq:Heq_galileon}
\end{eqnarray}
and
\begin{eqnarray}
0 &=&
\ddot{\phi} \bigg[ \alpha \dot{\phi}^2  - \left( \gamma + \frac{\alpha ^2}{\lambda^3} \right) \left(216  H^4 \sigma  \dot{\phi}^4 + 6 H \dot{\phi}^3 \right) + 6 \gamma H \dot{\phi}^3 + 108 H^4 M \sigma  \dot{\phi}^2+27 \zeta  H^2 \dot{\phi}^4 \bigg] \nonumber\\
& & + 3 \gamma \left(3 H^2+\dot{H}\right) \dot{\phi}^4 +3 \alpha  H \dot{\phi}^3 +9 \zeta  \left(3 H^3+2 H \dot{H} \right) \dot{\phi}^5 \nonumber\\
& & - \left( \gamma + \frac{ \alpha ^2 }{\lambda^3} \right) \left( 3  \left(3 H^2+\dot{H}\right) \dot{ \phi }^4 + 72  \sigma  \left(3 H^5+4 H^3 \dot{H}\right) \dot{ \phi }^5 \right) \nonumber\\
& & + 54 H^3 \sigma  \left(3 H^2+4 \dot{H}\right) \dot{\phi}^2 \left(\sqrt{2} g+2 M \dot{\phi}\right)+ \dot{\phi}^2 \left(3 H^2 M+\lambda ^3\right) \,.\label{eq:Seq_galileon}
\end{eqnarray}

A crucial observation is that the above field equations reduce to the ones of B$\flat$M \cite{Appleby:2020dko} in the limit $\left( \zeta, \sigma \right) \rightarrow 0$ and $\gamma \rightarrow - \alpha^2/\lambda^3$. This can be confirmed quite easily \footnote{See Eqs. (4.11-13) of Ref. \cite{Appleby:2020dko} (Fugue in B$\flat$).}. This information particularly points to a background equivalence between the Horndeski braiding Lagrangian $G(X) \sim M \ln(X/X_0)/2$ (Eq.~(\ref{eq:G_BFlatM})) and the Teledeski Lagrangian $\mathcal{G}(X, T, I_2) \sim I_2 \left( M + g/\sqrt{X} \right)$ (Eq.~(\ref{eq:mathcalG_galileon})) and adds further evidence into the existence of various Teledeski theories that may be tangent to the same Horndeski theory. Notably, this implies that the braiding and Teledeski Lagrangians may be transformed into each other. However, we stress that the braiding and the Teledeski potentials were motivated differently, i.e., the braiding Galileon appeared as a means of protection from radiative corrections while the Teledeski Lagrangian as a consequence of a softened Lovelock theorem in TG. A similar situation was also recognized in our earlier work wherein an infinite set of Teledeski cosmologies accommodating a well-tempered \textit{de Sitter} vacuum were found to be dynamically equivalent to a Horndeski theory \cite{Bernardo:2021izq}. We find that this recognition opens up the broader cosmological dynamics in TG. We move ahead working with the broader cases $\left( \zeta, \sigma \right) \neq 0$ and $\gamma \neq - \alpha^2/\lambda^3$.

We use the field equations (\ref{eq:Feq_galileon}), (\ref{eq:Heq_galileon}), and (\ref{eq:Seq_galileon}) in the next sections to look at the stability properties of the well-tempered Minkowski vacuum.

\subsection{Linear dynamical stability}
\label{subsec:linear_stability}

We take a look at the linearized perturbations $\left( \delta \phi,\, \delta H \right)$ on the well-tempered Minkowski vacuum
\begin{eqnarray}
H(t) &\rightarrow& \delta H(t)\,, \\
\phi(t) &\rightarrow& \phi(t) + \delta \phi (t) \,,
\end{eqnarray}
where $\phi$ on the right hand side is the on-shell scalar field. This is always a useful starting point in analyzing the dynamics in a self-tuning cosmology and most importantly it provides an objective way to determine when a numerical solution starts to fall towards the vacuum state.

By expanding the field equations about the perturbations $\left( \delta \phi, \delta H \right)$, we easily obtain the on-shell solution $\phi(t) = c_1 + c_2 t - t^2 \lambda^3 / \left( 2 \alpha \right)$ and the Hamiltonian constraint $\rho_\Lambda + \lambda^3 c_1 + \alpha c_2^2/2 = 0$ arising from the zeroth-order terms in the series. Furthermore, we obtain the linearized equations
\begin{equation}
\label{eq:FeqBbM7_linear}
0 = \frac{3 \alpha ^2  \dot{\phi}^3 \delta H }{\lambda ^3}-\alpha \dot{ \phi } \dot{ \delta \phi} - \lambda ^3 \delta \phi \,,
\end{equation}
\begin{equation}
0 = \delta H \left(6 M \dot{\phi} - \frac{9 \dot{\phi}^3 \left(\alpha ^2+2 \zeta  \lambda ^3 \ddot{\phi} \right)}{\lambda ^3}\right) + \dot{ \delta H } \left(6 +6 M \phi-\frac{1}{2} 9 \zeta  \dot{\phi}^4\right) +\frac{3 \alpha ^2   \dot{ \phi }^2 \ddot{\delta \phi}}{\lambda ^3} \,,\label{eq:HeqBbM7_linear}
\end{equation}
and
\begin{equation}
\label{eq:SeqBbM7_linear}
0 = \frac{3 \alpha ^2 \dot{\phi}^2 \dot{\delta H}}{\lambda ^3}+\frac{3 \alpha  \dot{\phi} \left(2 \alpha \ddot{ \phi }-\lambda ^3\right) \delta H }{\lambda ^3}-\alpha \ddot{ \delta \phi } \,.
\end{equation}
We can solve the Hubble perturbation $\delta H$ by eliminating $\delta \phi$ in Eqs.~(\ref{eq:HeqBbM7_linear}) and (\ref{eq:SeqBbM7_linear}). This leads to the asymptotic solution
\begin{equation}
\label{eq:deltaH_galileon}
\delta H \sim \frac{ \delta c_1 }{t^4} + O \left( t^{-5} \right) \,,
\end{equation}
where $\delta c_1$ is an integration constant. Substituting this into the constraint Eq.~(\ref{eq:FeqBbM7_linear}), we obtain
\begin{equation}
\label{eq:deltaphi_galileon}
\delta \phi \sim \lambda^3 \delta c_2 t - \alpha c_2 \delta c_2 + O \left( t^{-1} \right)\,, 
\end{equation}
where $\delta c_2$ is an integration constant.

The important observation is that the Hubble perturbation vanishes and that the scalar perturbation $\delta \phi$ can be absorbed into the integration constants of $\phi$ on-shell. This shows the stability of the well-tempered Minkowski vacuum under linear perturbations. Later, we make good use of the linear asymptotic solutions (\ref{eq:deltaH_galileon}) and (\ref{eq:deltaphi_galileon}) to assess when the state is settling down the Minkowski vacuum. We proceed in the next section to look at the dynamics beyond a linear analysis by drawing the vector fields of the dynamical system.

\subsection{Phase portraits: A look at the global dynamics}
\label{subsec:phase_portraits}

In this section, we look at the cosmological dynamics beyond the linear analysis by making use of a dynamical system analysis. This popular approach is an illustrative way to gain insight on a system particularly when an analytical solution is unavailable \cite{Bahamonde:2017ize}.

We can obtain a two-dimensional dynamical system with the modified Friedmann equations (\ref{eq:Feq_galileon}) and (\ref{eq:Heq_galileon}) and the scalar field equation (\ref{eq:Seq_galileon}) with vacuum energy, i.e., $P_\Lambda = - \rho_\Lambda$. In particular, we eliminate the explicit $\phi(t)$ dependence in the Hubble equation (\ref{eq:Heq_galileon}) by using the Friedmann constraint (\ref{eq:Feq_galileon}). The resulting equation together with the scalar field equation then make up a dynamical system for the vector field $( X , H )$ where $X = \dot{\phi}^2/2$ is the kinetic term. In addition, we use the tadpole coefficient $\lambda$ to define a length scale $L = 1/\lambda^{3/2}$ through which dimensionless versions of the theory parameters and the fields can be defined. A straightforward dimensional analysis leads to the transformation $H(\tau) = y(\tau)/L$, $\dot{\phi}^2/2 = \chi/L^2$, $\rho_\Lambda = l^2/L^2$, $\gamma = L^2 \bar{\gamma}$, $\zeta = L^4 \bar{\zeta}$, $g = \bar{g}/L$, and $\sigma = L^4 \bar{\sigma}$. We note that the parameters $\alpha$ and $M$ are already dimensionless in geometrized units. In what follows, for brevity, we omit the bars over the dimensionless parameters $(\bar{\gamma}, \bar{\zeta}, \bar{g}, \bar{\sigma})$, write them as simply $(\gamma, \zeta, g, \sigma)$, and put up a warning should a distinction is necessary. The dynamical system then becomes explicitly expressed in terms of the dimensionless fields $\left( \chi (\tau), y(\tau) \right)$ where $\tau = t/L$ is a dimensionless time variable.

Taking the steps mentioned in the previous paragraph, we obtain the two-dimensional dynamical system:
\begin{equation}
\label{eq:chip_galileon}
    \chi' = - \frac{Q[\chi, y]}{U[\chi, y]}\,,
\end{equation}
and
\begin{equation}
\label{eq:yp_galileon}
    y' = - y \left(3 M y^2-1\right) \frac{S[\chi, y]}{W[\chi, y]} \,,
\end{equation}
where a prime in $\chi\left(\tau\right)$, $y\left(\tau\right)$ denotes a derivative with respect to the dimensionless time $\tau$, and the functionals $Q$, $S$, $U$, and $V$ are explicitly provided in \ref{sec:UW_galileon}. Obviously, the above dynamical system looks repulsive and even more so when the functionals are written in full. Nonetheless, it can be confirmed that in the Minkowski limit ($y(\tau) \rightarrow 0$) they reduce to $\chi' = - \sqrt{2\chi}/\alpha$ and $y' = 0$ which can be easily recognized as describing the on-shell solution $\phi(\tau) = c_1 + c_2 \tau - \tau^2/(2\alpha)$.

\begin{figure}[h!]
\center
	\subfigure[ B$\flat$m $\left(M = 0\right)$ ]{
		\includegraphics[width = 0.45 \textwidth]{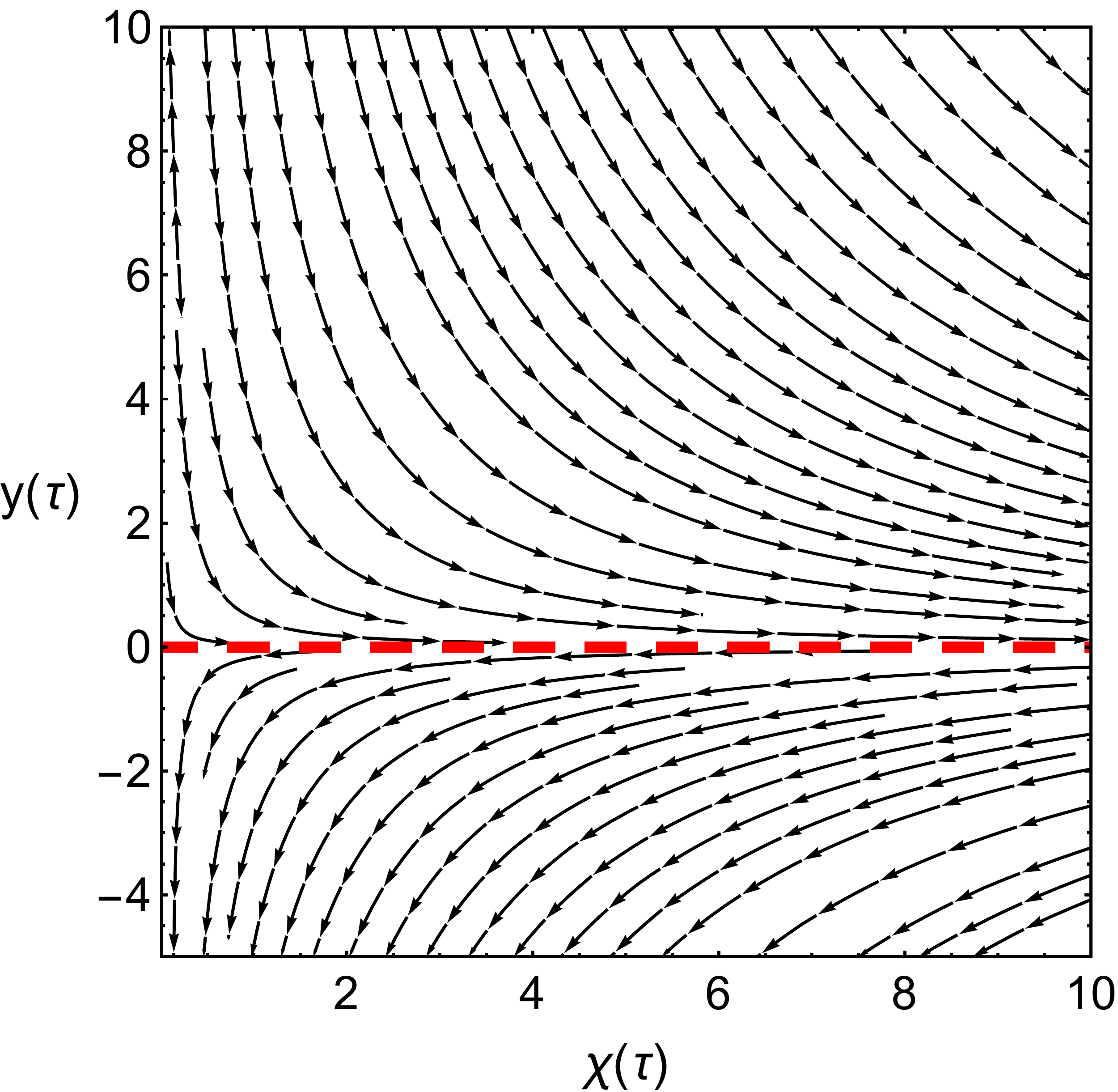}
		}
	\subfigure[ B$\flat$M $\left(M = 10^{-8}\right)$ ]{
		\includegraphics[width = 0.45 \textwidth]{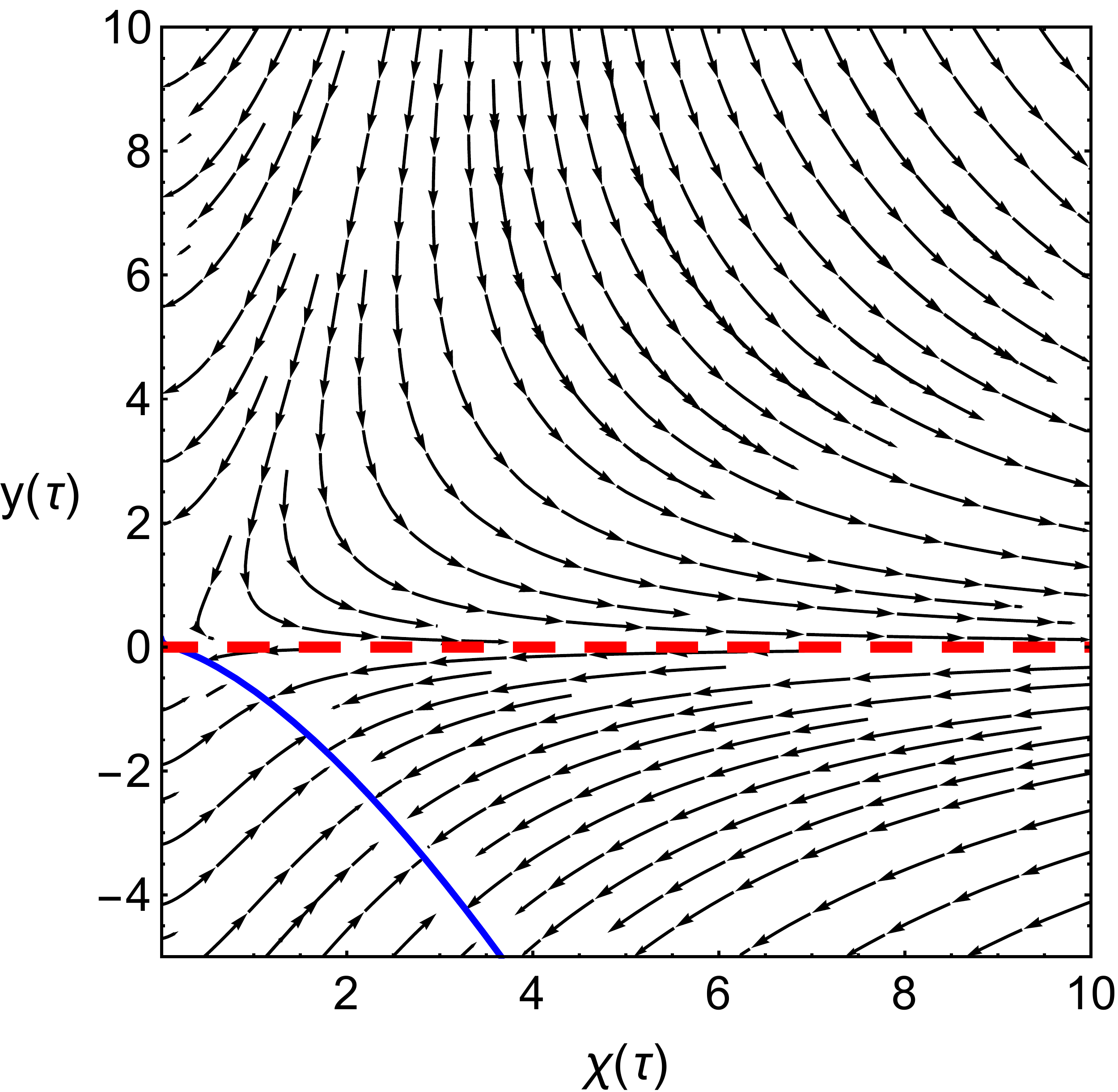}
		}
\caption{Phase portraits of the (a) B$\flat$m and (b) B$\flat$M systems for a vacuum energy of $\rho_\Lambda = 10^{10}/L^2$ and theory parameter $\alpha = 10$ where $L$ is the tadpole length scale. $y$ and $\chi$ are the dimensionless Hubble function ($H(\tau) = y(\tau)/L$) and kinetic term ($\dot{\phi}^2/2 = \chi/L^2$), respectively. The red-dashed line is the Minkowski vacuum and the blue-solid line is the place where the dynamical system becomes undefined.}
\label{fig:Bb}
\end{figure}

We present the phase portraits of the dynamical system (Eqs.~(\ref{eq:chip_galileon}) and (\ref{eq:yp_galileon})). To establish a baseline, we first look at the B$\flat$ phase portraits and then draw the ones that are influenced by the inclusion of TG. The phase portraits of the B$\flat$m and B$\flat$M are shown in Fig.~\ref{fig:Bb}.

In Fig.~\ref{fig:Bb}, we consider a vacuum energy of size $\rho_\Lambda \sim 10^{10}/L^2$ and fixed $\alpha \sim 10$ where $L$ is the tadpole length scale. The physical region is understandably $y = H L > 0$ and for this Fig.~\ref{fig:Bb} shows that the system would eventually end up on-shell or rather fall to the well-tempered Minkowski vacuum state. It is also important to point out that for $y < 0$ the phase portraits show that the system may end up somewhere other than the Minkowski vacuum. This suggests an apparent instability because the dynamical system behaves differently for $y > 0$ and $y < 0$. We suspect this feature enters through the braiding and the Teledeski terms which comes with terms usually coupled to odd powers of the Hubble function. Nonetheless, the $y < 0$ is harmless as long as it is kinematically inaccessible. This was shown to be the case for linear perturbations (Sec.~\ref{subsec:linear_stability}) which do not depend on the sign of the $\delta H$. Also, in practice, it is worth noting that only the region $y > 0$ can be considered to be physically relevant and Fig.~\ref{fig:Bb} shows that the dynamical system inevitably end up on-shell in this phase space. We found no evidence in our various numerical analyses that the nonphysical region $y < 0$ is kinematically accessible to the physical region $y > 0$. As a matter of fact, it can be shown using the field equations (\ref{eq:Feq_galileon}-\ref{eq:Seq_galileon}) that $H = 0$ implies $\dot{H} = 0$, i.e., there is no gradient in the $H$-direction in the vacuum state. The stability of the well-tempered Minkowski vacuum can be viewed from this standpoint.

Fig.~\ref{fig:Bb} also shows the effect of a conformal coupling $M \phi$ to the Ricci scalar. Clearly, in Fig.~\ref{fig:Bb}(a), the stream lines in the physical region $y > 0$ can be easily argued to be consistently approaching the Minkowski vacuum. However, this changes in Fig.~\ref{fig:Bb}(b) ($M \neq 0$) where a region of the phase space (upper left) of the physically relevant $y > 0$ can now be seen to approach the vertical line $\chi = 0$ rather than fall to the Minkowski vacuum. This behavior can be understood through an asymptotic solution with $M \neq 0$ for $\chi \ll 1$: $y(\chi) \sim - \left( C + \sqrt{C^2 + 12 M} \right) / (6M)$ where $C$ is an integration constant. This confirms that with $M \neq 0$ (B$\flat$M) the stream lines for $\chi \ll 1$ can approach the vertical line $\chi = 0$ at some constant $y$. In addition to this, in the unphysical region $y < 0$, Fig.~\ref{fig:Bb}(b) presents further a place (blue-solid line) where the dynamical system becomes undefined. We note that this is a harmless, generic feature in phase portraits and can be traced to the choice of the coordinates in phase space. In Fig.~\ref{fig:Bb}(b), it also persists only in the unphysical $y < 0$ but it is nonetheless a feature with $M \neq 0$ worth pointing out.

\begin{figure}[ht]
\center
	\subfigure[ B$\flat$m Extension: $\left(M = 0, \sigma > 0\right)$ ]{
		\includegraphics[width = 0.45 \textwidth]{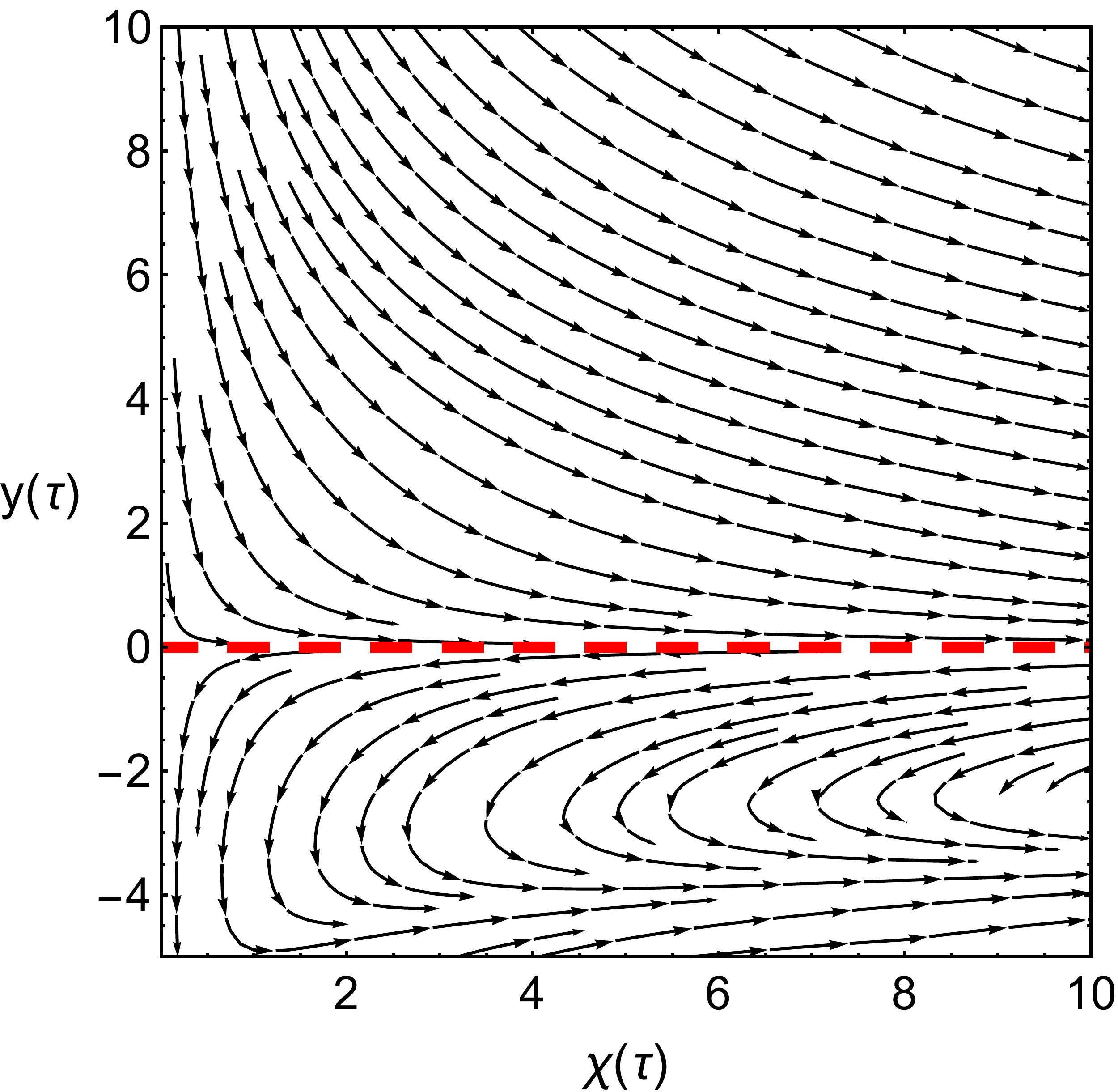}
		}
	\subfigure[ B$\flat$M Extension $\left(M = 10^{-8}, \sigma > 0\right)$ ]{
		\includegraphics[width = 0.45 \textwidth]{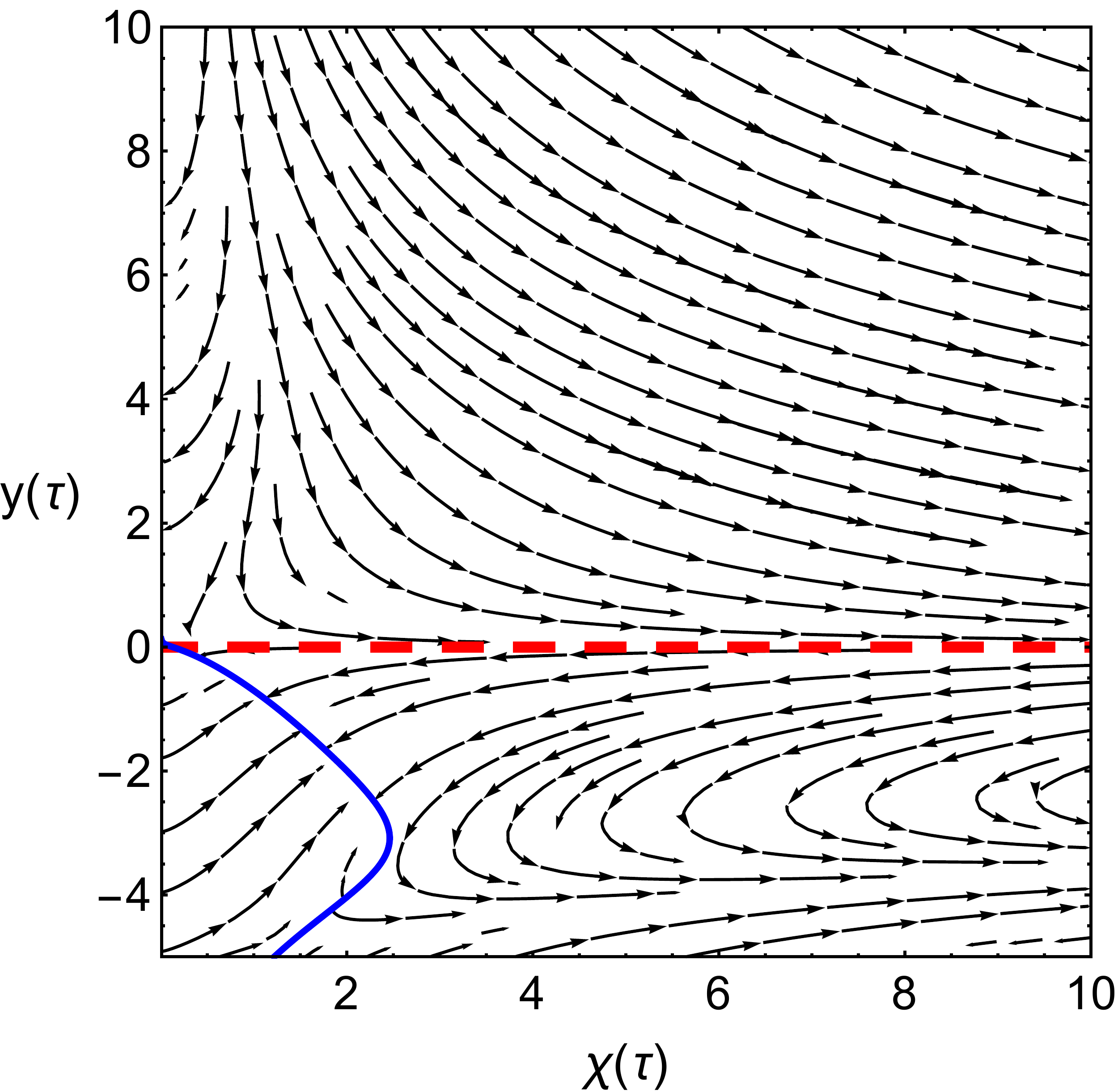}
		}
	\subfigure[ B$\flat$m Extension $\left(M = 0, \sigma < 0\right)$ ]{
		\includegraphics[width = 0.45 \textwidth]{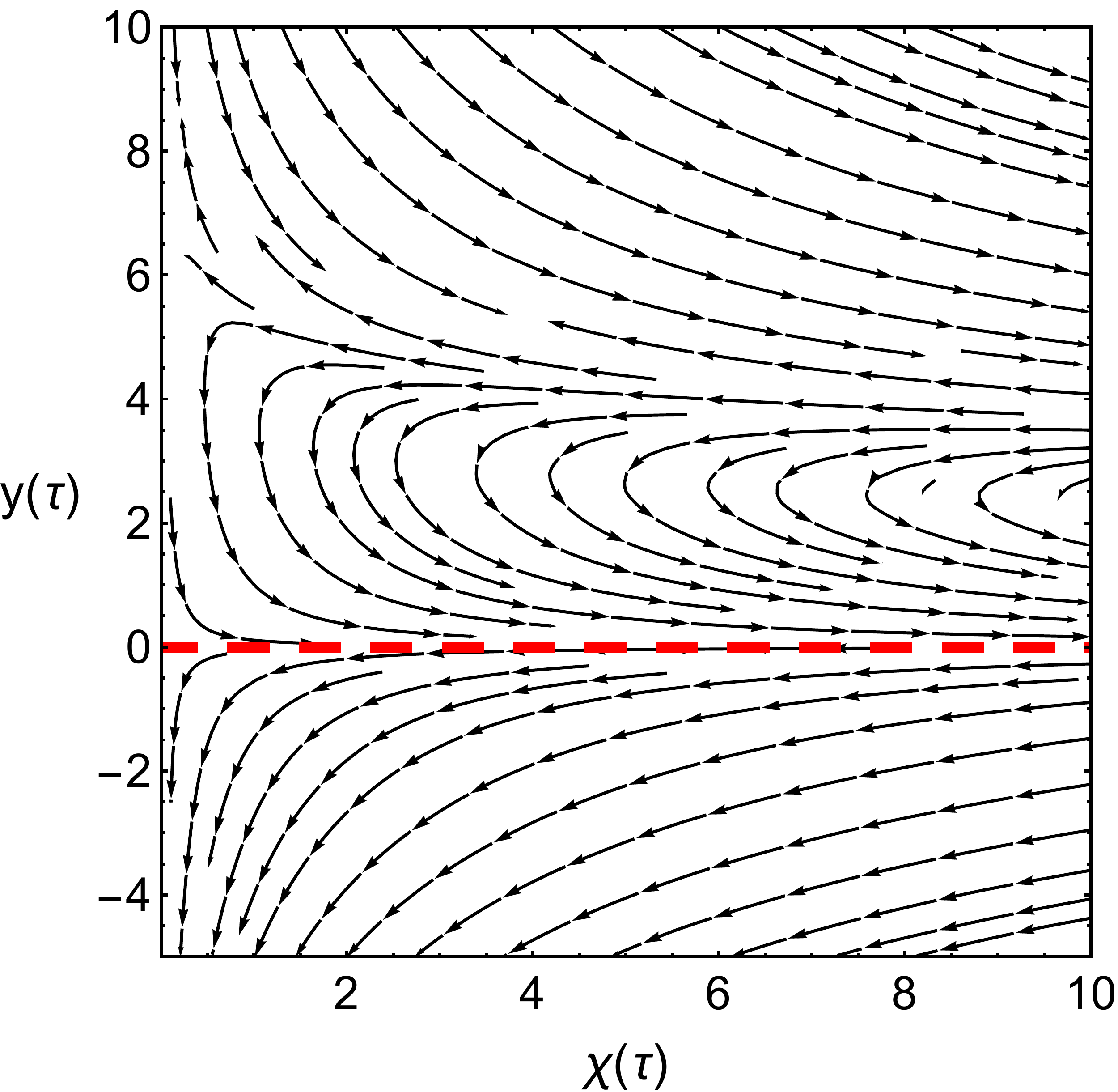}
		}
	\subfigure[ B$\flat$M Extension $\left(M = 10^{-8}, \sigma < 0 \right)$ ]{
		\includegraphics[width = 0.45 \textwidth]{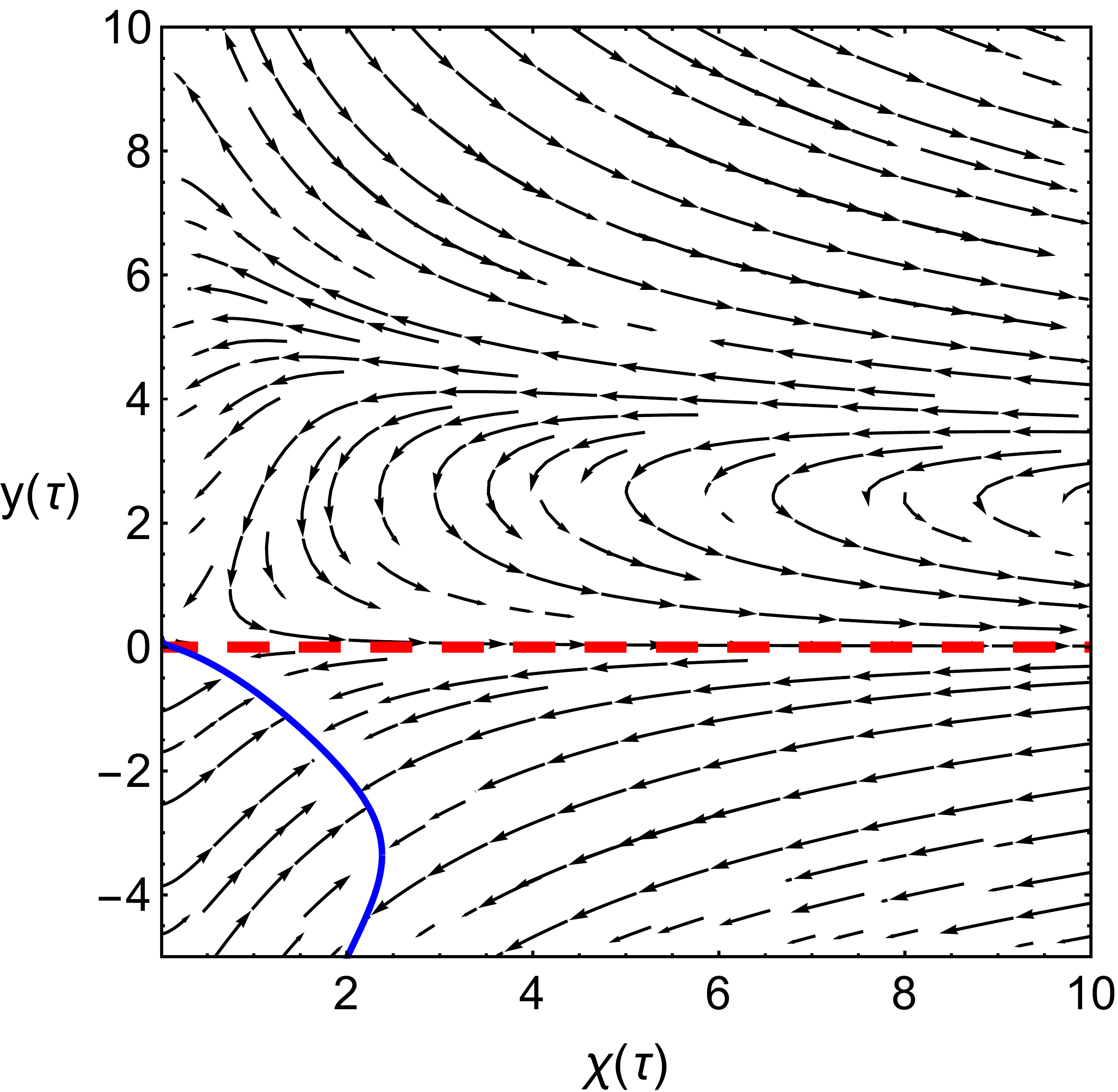}
		}
\caption{Phase portraits of the well-tempered teleparallel Galileon for a vacuum energy of $\rho_\Lambda = 10^{10}/L^2$ and fixed theory parameters $\alpha = 10$, $\gamma = - L^2$, $\zeta = 0$, $g = 0$, and $\sigma = \pm 10^{-4} L^4$ where $L$ is the tadpole length scale. $y$ and $\chi$ are the dimensionless Hubble function ($H(\tau) = y(\tau)/L$) and kinetic term ($\dot{\phi}^2/2 = \chi/L^2$), respectively. The red-dashed line is the Minkowski vacuum and the blue solid line is the place where the dynamical system becomes undefined.}
\label{fig:wt_tg_galileon}
\end{figure}

Now, we move to the phase portraits of the dynamical system (\ref{eq:chip_galileon}) and (\ref{eq:yp_galileon}) with the Teledeski extension ($\sigma \neq 0$). These are presented in Fig.~\ref{fig:wt_tg_galileon}. 

In Fig.~\ref{fig:wt_tg_galileon}(a) it can be seen that with $M = 0$ and $\sigma > 0$ the stream lines in the physical region ($y > 0$) show support to the well-tempered Minkowski vacuum (red-dashed line) as a late-time state. This is also consistent with the B$\flat$m (Fig.~\ref{fig:Bb}(a)) wherein all of the stream lines above the horizontal line $y = 0$ point to the Minkowski vacuum. However, it is more interesting to point out the drastic change in the picture which happened in the nonphysical region $y < 0$. This time ($\sigma > 0$) the stream lines in $y < 0$ can be seen to be exhibiting a turning point at some low-$\chi$ threshold. Now, in Fig.~\ref{fig:wt_tg_galileon}(b) with $M > 0$ and $\sigma > 0$ it is also the behavior for $y < 0$ that was largely affected with the inclusion of $\sigma$. A similar low-$\chi$ turning point threshold also appeared in addition to a change in the shape of the blue-solid curve compared with the one with $\sigma = 0$ (Fig.~\ref{fig:Bb}(b)). It gets more interesting to see the $\sigma < 0$ phase portraits. In Figs. \ref{fig:wt_tg_galileon}(c-d), it can be seen that the low-$\chi$ turning point threshold now occurred above the horizontal line $y = 0$ in the physical region. This can be traced to the fact that the Teledeski terms in the dynamical system (Eqs.~(\ref{eq:chip_galileon}) and (\ref{eq:yp_galileon})) that $\sigma$ parametrizes are mostly coupled to odd powers of the Hubble function. Therefore, in Figs. \ref{fig:wt_tg_galileon}(c-d) we can see that the low-$\chi$ turning point threshold influenced the physically relevant region $y > 0$. This is an important point and it adds huge value to the inclusion of the Teledeski terms that vanish on-shell but can affect the dynamics off-shell in a multitude of ways. Clearly, with $\sigma < 0$, the stream lines in Figs. \ref{fig:wt_tg_galileon}(c-d) show that the dynamics in a portion of the physical region $y > 0$ continues to end up on-shell. However, the stream lines above some $y = y_{+} > 0$ for $\sigma < 0$ do not any more seem to suggest that the dynamics could end up on the well-tempered vacuum. This can be observed in particular for the vector fields in Figs. \ref{fig:wt_tg_galileon}(c-d) above $y \sim 4$. Therefore, the inclusion of $\sigma$ does change the dynamics drastically and reveals a glimpse of dynamics that can be obtained in TG.

This is a good place to remind ourselves that the term $\sigma T I_2^2$ in the Teledeski sector of the well-tempered action (\ref{eq:action_wt_galileon}) is just one simple choice among a family of analytic functions $F$ (Eq.~(\ref{eq:FugueBFlatM7})) that would leave the on-shell conditions unaffected. In practice, the function $F$ will generate interesting off-shell cosmological dynamics that could potentially be constrained with Hubble data. This attests to the richer cosmology accessible in TG.

\subsection{Response to a Phase Transition}
\label{subsec:phase_transition}

In this section, we numerically explore a scenario in which the vacuum energy transitions from an initial lower value to a final one -- a \textit{phase transition} -- to assess whether the well-tempered vacuum continues to be an attractor under such a condition.

\begin{figure}[ht]
\center
\includegraphics[width = 0.45 \textwidth]{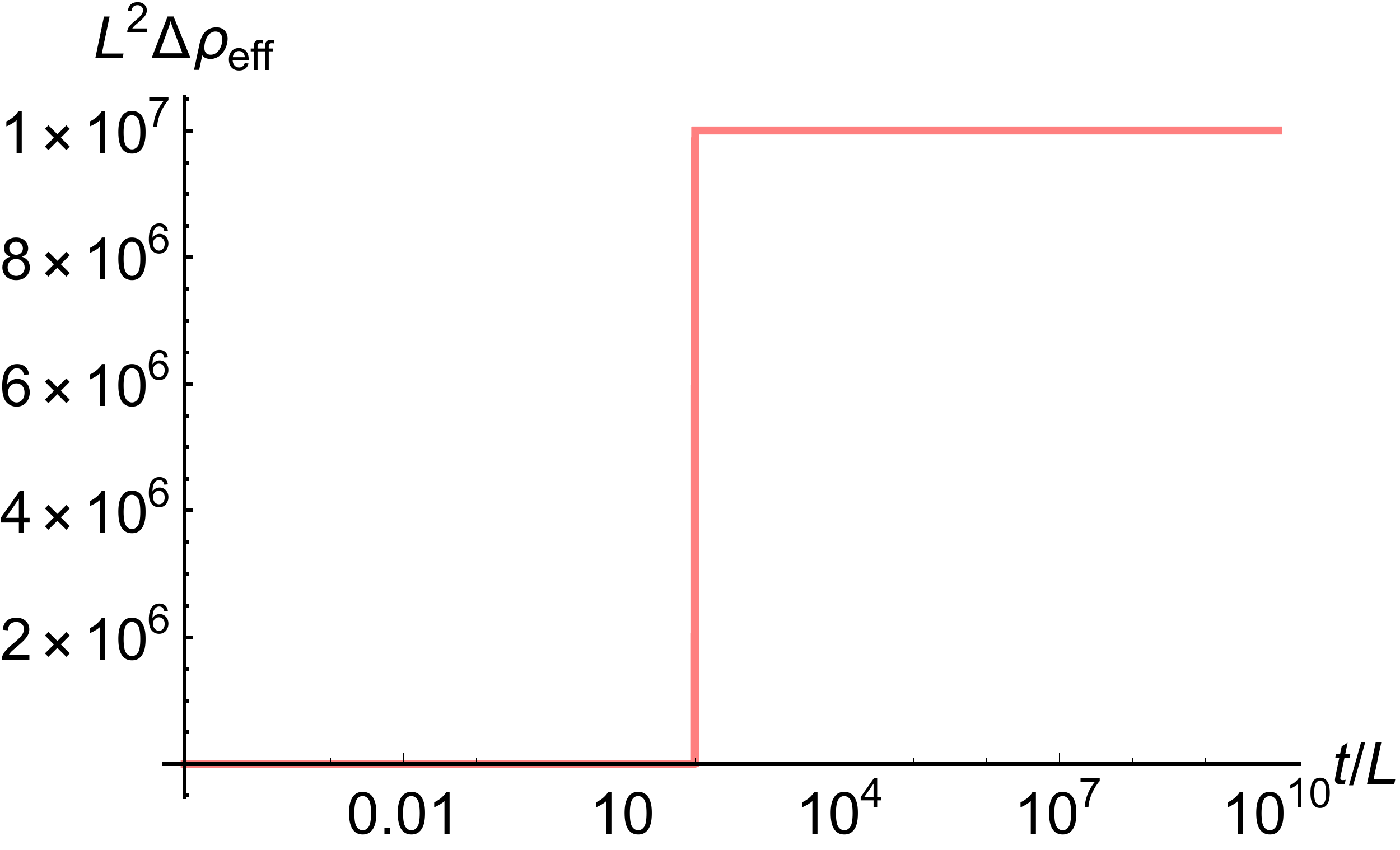}
\caption{A vacuum energy $\rho_\Lambda$ in a phase transition of magnitude $\Delta \rho_\Lambda = 10^7 / L^2$ at a time $t = 10^2 L$ for an interval $\Delta t = L/10$ where $L$ is the tadpole length scale.}
\label{fig:PT}
\end{figure}

We implement the phase transition through an effective energy density given by Refs.~\cite{Emond:2018fvv, Bernardo:2021hrz,  Bernardo:2021izq}
\begin{equation}
\label{eq:rho_eff}
\rho_{\rm{eff}} = \rho_\Lambda + \frac{ \Delta \rho_\Lambda }{2} \tanh \left( \frac{t - T}{\Delta T} \right)\,,
\end{equation}
This is clearly describing a vacuum energy in transition from an initial energy density $\rho_\Lambda - \Delta \rho_\Lambda/2$ ($t \ll T$) to a final value $\rho + \Delta \rho_\Lambda/2$ ($t \gg T$) in an interval $\Delta T$ at the time $t \sim T$. This is shown in Fig.~\ref{fig:PT} where $T \sim 10^2 L$, $\Delta T \sim L/10$, and $\Delta \rho_\Lambda \sim 10^7 / L^2$ where $L$ is the tadpole length scale. In the limit of $\Delta T$ being much smaller than the time scale of the evolution, the effective energy density (\ref{eq:rho_eff}) can be viewed as merely a step function from the initial vacuum energy to the final one, as it appears in Fig.~\ref{fig:PT}.

We input the effective source (Eq. (\ref{eq:rho_eff})) in the modified Friedmann and scalar field equations and then numerically integrate to see if the system may accommodate a phase transition all while falling to the Minkowski vacuum. Using the tadpole as a reference length scale $L$, we define the dimensionless theory parameters $\left( \bar{\gamma}, \bar{\zeta}, \bar{g}, \bar{\sigma} \right)$ by $\gamma = L^2 \bar{\gamma}$, $\zeta = L^4 \bar{\zeta}$, $g = \bar{g}/L$, $\sigma = L^4 \bar{\sigma}$ and dimensionless phase transition parameters $\left( l, \delta, \bar{T}, \bar{\Delta T} \right)$ by $\rho_\Lambda = l^2/L^2$, $\Delta \rho_\Lambda = l^2 \Delta / L^2$, $T = \bar{T} L$, and $\Delta T = \bar{\Delta T} L$. We then choose the phase space variables of the system to be $H_0$ and $\dot{\phi}_0$ where a subscript ``$0$'' refers to the dynamical variables being evaluated at an initial time. The numerical integration is then carried forward in time fixing $\phi_0$ initially in the Hamiltonian constraint. We make sure that the Hamiltonian constraint is satisfied throughout the evolution.

\begin{figure}[ht]
\center
	\subfigure[ ]{
		\includegraphics[width = 0.45 \textwidth]{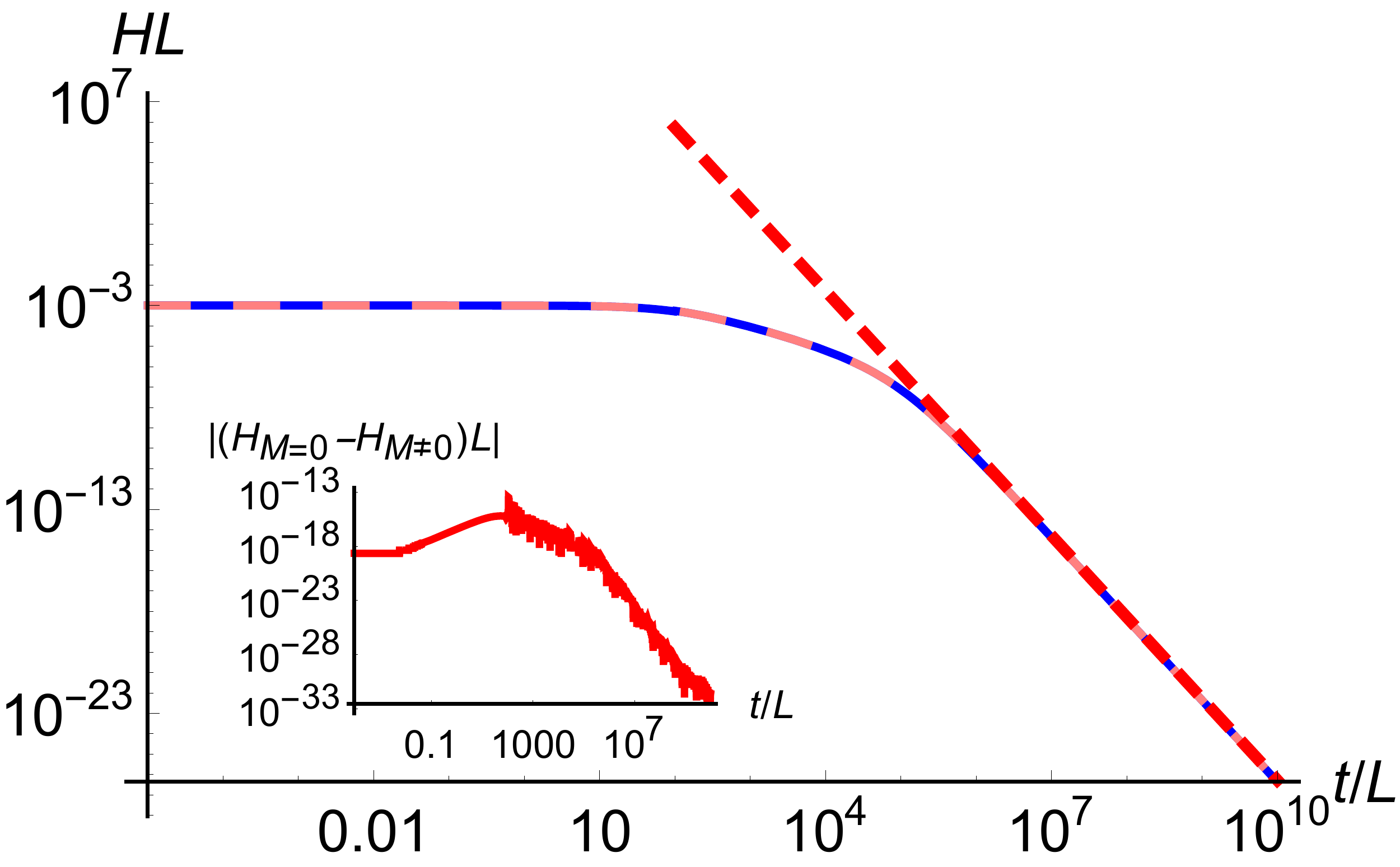}
		}
	\subfigure[ ]{
		\includegraphics[width = 0.45 \textwidth]{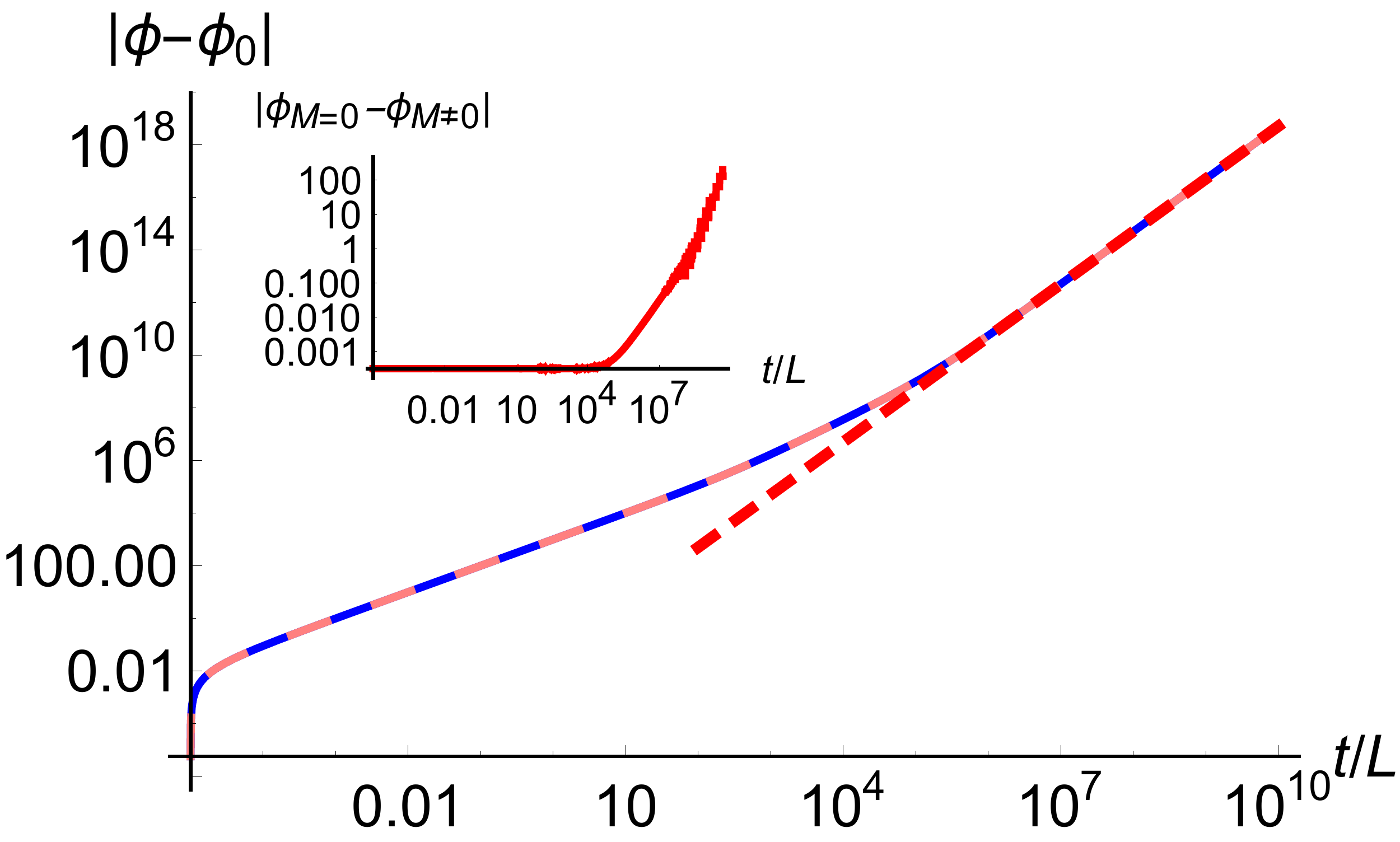}
		}
\caption{The (a) Hubble function and (b) scalar field solution to the well-tempered teleparallel Galileon immersed an initial vacuum energy $\rho_\Lambda = 10^{10}/L^2$ in a phase transition of magnitude $\Delta \rho_\Lambda = 10^7 / L^2$ at a time $t = 10^2 L$ for an interval $\Delta t = L/10$ where $L$ is the tadpole length scale. The theory parameters are fixed to $\alpha = 10$, $\gamma = - L^2$, $\zeta = 0$, $g = 0$, and $\sigma = 10^{-4} L^4$. The blue-solid line corresponds to $M = 0$ and and the pink-long-dashed line to $M = 10^{-8}$. The red-short-dashed line are the on-shell asymptotic solutions (\ref{eq:deltaH_galileon}) and (\ref{eq:deltaphi_galileon}).}
\label{fig:wt_tg_galileon_PT}
\end{figure}

Fig.~\ref{fig:wt_tg_galileon_PT} shows a result of the numerical integration with the phase transition in Fig.~\ref{fig:PT} in both the B$\flat$m and B$\flat$M extensions.

We find that as long as $H_0 > 0$, the system inevitably ends up on-shell at late times, echoing the dynamics displayed in the previous section in terms of phase portraits. Furthermore, we find no evidence of the system beginning from $H_0 > 0$ and eventually falling to the nonphysical region $H_0 < 0$ provided that the vacuum energy transition is positive. This is of course reasonable as phase transitions of the vacuum energy are expected to be positive in a particle physics perspective. Thus, our numerical results support the assertion that the nonphysical region $H_0 < 0$ is kinematically inaccessible. From this standpoint, we view the stability of the well-tempered Minkowski vacuum.

\begin{figure}[ht]
\center
\includegraphics[width = 0.45 \textwidth]{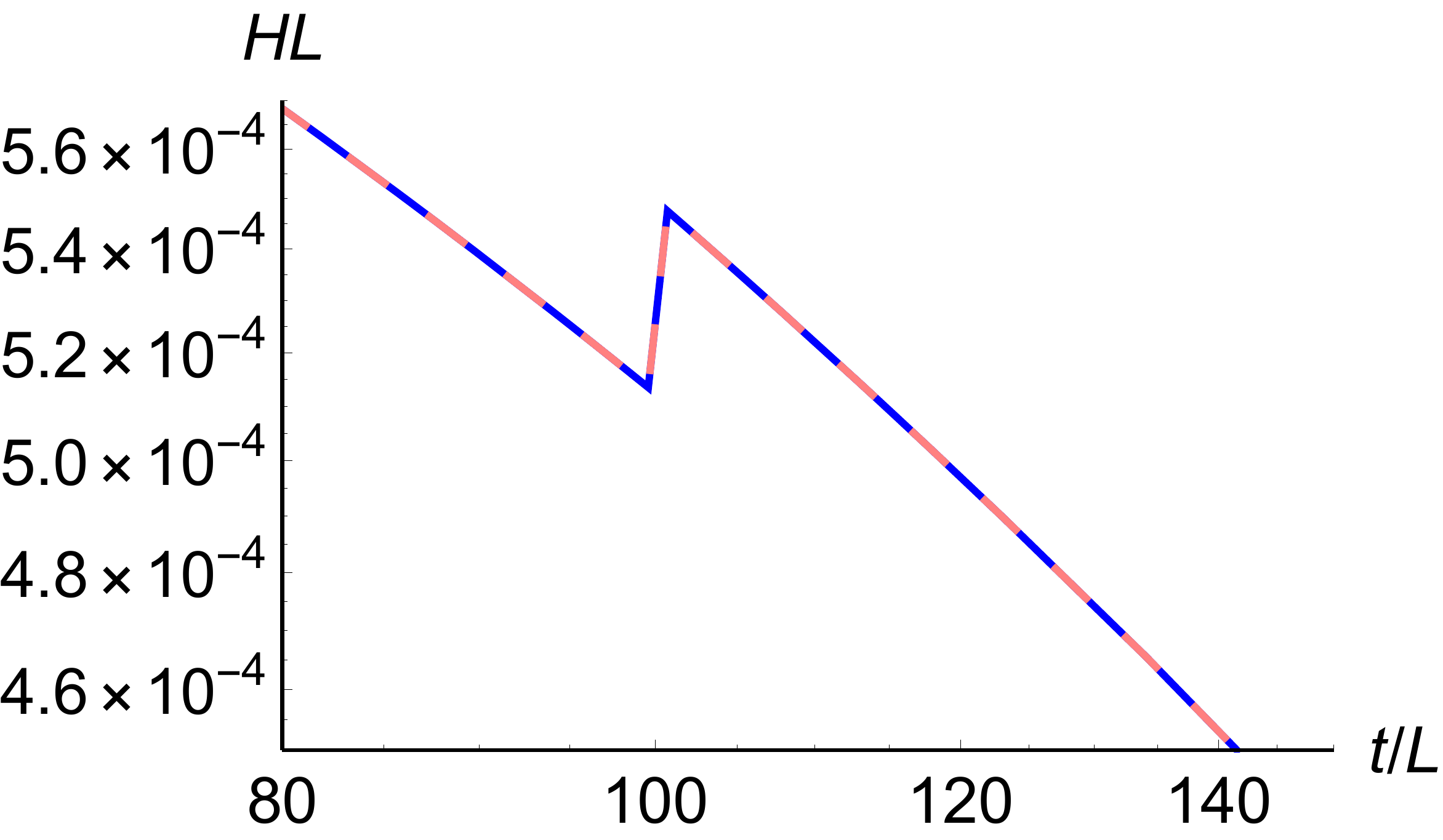}
\caption{A magnified view of the Hubble function in Fig.~\ref{fig:wt_tg_galileon_PT}(a) responding to a positive phase transition of magnitude $\Delta \rho_\Lambda \sim 10^{7}/L^2$ shown in Fig.~\ref{fig:PT}.}
\label{fig:wt_tg_galileon_PT_zoomin}
\end{figure}

Fig.~\ref{fig:wt_tg_galileon_PT} reflects this discussion with a phase transition of the size $\Delta \rho_\Lambda \sim 10^{7}/L^2$ (in words, \textit{seven} orders of magnitude above the tadpole's energy scale). In Fig.~\ref{fig:wt_tg_galileon_PT}(a), the Hubble function can be seen to be initially loitering in an initial quasi de Sitter phase but then eventually at late times the solution approaches the well-tempered Minkowski vacuum (red-short-dashed line). The red-short-dashed lines in Fig.~\ref{fig:wt_tg_galileon_PT} are given by the on-shell asymptotic solutions (\ref{eq:deltaH_galileon}) and (\ref{eq:deltaphi_galileon}) which provide an objective way to determine when the system begins to fall to the Minkowski vacuum. It is also worth noting that the visual-indistinguishability between the $M = 0$ and $M \neq 0$ trajectories in Fig.~\ref{fig:wt_tg_galileon_PT} is due to the same initial conditions being in use in both cases and the fact that the on-shell behavior is independent of $M$. The inset was included to highlight the numerical resolvable difference between the $M = 0$ and $M \neq 0$ curves. We must further point out that the Hubble function responded in the expected way to a positive phase transition. However, this cannot be seen in Fig.~\ref{fig:wt_tg_galileon_PT} due to the logarithmic scale of the axes. A zoomed-in view of the Hubble function responding to the phase transition is shown in Fig.~\ref{fig:wt_tg_galileon_PT_zoomin}.

A positive phase transition corresponds to a more negative pressure from the vacuum energy and so arguably leads to an increase in the expansion rate which translates to a positive increase in the Hubble function. This description is captured in Fig.~\ref{fig:wt_tg_galileon_PT_zoomin}.

The titanic difference in the scales of Figs.~\ref{fig:wt_tg_galileon_PT} and \ref{fig:wt_tg_galileon_PT_zoomin} also shows that the dynamics spends most of its time loitering in a quasi de Sitter state instead of immediately falling down to the Minkowski vacuum. We find these loitering phases to be quite persistent in our simulations. This can possibly be traced to the fact that no mass scale can be associated with a Minkowski vacuum. In our numerical analysis, we considered the tadpole length as a reference length scale although any one of the dimensionfull parameters in the action may be arbitrarily alternatively chosen. This is an important physical difference between models being endowed with Minkowski or de Sitter solutions. Obviously, de Sitter states come with a natural mass scale, i.e., the Hubble parameter, that also consequently defines the time scales, while Minkowski solutions do not.

\section{Conclusion}
\label{sec:conclusions}

The standard model of cosmology sets aside the late-time, low energy cosmic acceleration from the enormous, possibly Planck scale, vacuum energy expected from particle physics. This cosmological constant problem together with the absence of a fundamental description of the dark components in the standard model has prompted the investigation of dynamical dark energy. It has been recognized that dark energy itself may be able to dynamically tune away vacuum energy. This thought process eventually led to the well-tempered models showcasing a low energy attractor vacuum despite the presence of a titanic cosmological constant.

We built on the Horndeski models known as Fugue in B$\flat$ \cite{Appleby:2020dko} to obtain a larger family of scalar-tensor models admitting a well-tempered Minkowski vacuum in the teleparallel Horndeski theory. This expansion was first and foremost possible due to the wiggle room in the Lovelock theorem afforded by TG leading to new, additional Horndeski terms in the action and inevitably much richer cosmological dynamics.

We setup the most general recipe for obtaining these extended Fugue in B$\flat$ models (Sec.~\ref{sec:well_tempered_recipe}) and presented a variety of examples (Sec.~\ref{sec:well_tempered_minkowski}), all of which are able to dynamically cancel out an arbitrarily large vacuum energy leaving only a Minkowski state. We further established that provided there is no explicit $\phi$ dependence in the action beyond a tadpole and a conformal coupling with the Ricci scalar, well-tempering is the only method of utilizing degeneracy in field space to cancel out the vacuum energy and retain a flat Minkowski state (Sec.~\ref{sec:trivial_scalar_minkowski}).

Through the well-tempered teleparallel Galileon (Sec.~\ref{sec:wt_galileon}), we have also demonstrated the persistence of the attractor nature of the Minkowski vacuum by means of linear stability analysis, phase portraits, and phase transitions.

We highlight the significance of the new quartic Horndeski and Teledeski couplings which leave the on-shell behavior unchanged but drastically influences the dynamics once the system is off-shell. These couplings remain unconstrained as far as well-tempering is concerned and so should be recognized as a profound addition to the theory that can potentially be constrained through observations.

Needless to say, the Universe today is undergoing cosmic acceleration. Nonetheless, the existence of a Minkowski vacuum, especially one which dynamically cancels out vacuum energy, is a reassuring feature in any theory, may it be designed for cosmology or elsewhere. Well-tempered Minkowski solutions may also be useful in a variety of applications outside of the cosmological context where the spacetime can be considered to a good extent as flat rather than curved.

We mention some analyses that remain to be addressed. First, the soundness of the theory in terms of the absence of ghost and gradient instabilities must be further evaluated. This is an important issue that hangs on the scalar perturbations of Teledeski cosmology which continue to be technically challenging due to the multitude of scalar couplings that can be hosted in TG. We paved the road by setting up the general recipe which can be subsequently supplemented with additional theoretical constraints. Observational constraints on well-tempered cosmology remain unexplored, regardless of the vacuum being of de Sitter or Minkowski type. We expect to fill in this gap in the literature in later work. Finally, we neglected the quintic Horndeski sector. It remains to be seen whether this sector influences the dynamics on-shell or if it enters only as an arbitrary coupling that disrupts the off-shell evolution.

\section*{Acknowledgments}\label{sec:acknowledgements} 
The authors thank Ian Vega for helpful discussion on dynamical systems. JLS would like to acknowledge networking support by the COST Action CA18108 and funding support from Cosmology@MALTA which is supported by the University of Malta. JLS would also like to acknowledge funding from ``The Malta Council for Science and Technology'' in project IPAS-2020-007. SA is supported by an appointment to the JRG Program at the APCTP through the Science and Technology Promotion Fund and Lottery Fund of the Korean Government, and was also supported by the Korean Local Governments in Gyeongsangbuk-do Province and Pohang City. MC would like to acknowledge funding by the Tertiary Education Scholarship Scheme (TESS, Malta).

\appendix

\section{Non-Galileon Well-Tempered Models}
\label{sec:non_galileon_models}

We present various illustrative examples of solving the degeneracy equation (\ref{eq:BFlatM7_recast}) to obtain well-tempered models with non-Galileon ansatzes to the Horndeski potentials.

\subsubsection*{B$\flat$TG ($\mathcal{G}$-extension of B$\flat$M)}

Suppose we put in the B$\flat$M model (Eqs.~(\ref{eq:V_BFlatM}) and (\ref{eq:G_BFlatM})) as an ansatz to the B$\flat$TG degeneracy equation (\ref{eq:BFlatM7_recast}). This seems a reasonable starting point. We find that this leads to
\begin{equation}
\tilde{\mathcal{G}}(X) = \frac{g}{\sqrt{X}}\,,
\end{equation}
where $g$ is a constant. The Teledeski potential then becomes $\mathcal{G}:= (g/\sqrt{X}) (I_2 + F(T, I_2))$ where $F(T, I_2) = \mathcal{O}(T, I_2^2)$. The Hamiltonian constraint on-shell can be shown to be $\rho_\Lambda + \lambda^3 c_2 = 0$ where $\phi(t) \sim c_2$ revealing the dynamical cancellation of the vacuum energy that is taking place.

\subsubsection*{B$\flat$TG ($V$-B$\flat$M only)}

We consider the general power law $V$ potential as in Eq.~(\ref{eq:V_BFlatM}) in B$\flat$M. This fully specifies the on-shell scalar field $\dot{\phi} = \left( - \left( 2n\lambda^3 t/\epsilon \right) + c_1 \right)^{1/(2n+1)}$ and the Hamiltonian constraint $\rho_\Lambda + \lambda^3 c_2 = 0$ where $\phi(t) \sim c_2$. However, this time, the degeneracy equation reduces to
\begin{equation}
\label{eq:BFlatM7T2}
6 (2 n+1) \epsilon^2 X^{2 n+1}=3 \lambda ^3 \left(-2 X G_X+M-2 X \tilde{\mathcal{G}}_X-\tilde{\mathcal{G}}\right) \,,
\end{equation}
implying that either $G$ or $\tilde{\mathcal{G}}$ have to be further specified in order to close the system. Since Eq.~(\ref{eq:BFlatM7T2}) is a linear differential equation in both $G$ or $\tilde{\mathcal{G}}$, we can choose to specify $G$ without loss of generality. In this direction, we consider a power law braiding $G$ of the form
\begin{equation}
G(X) = - \frac{ \eta \epsilon^2 X^{1 + 2 m} }{ \lambda^3 }\,,
\end{equation}
where $\eta$ and $m$ are constants. This form is motivated by the first term of Eq.~(\ref{eq:G_BFlatM}) but with a different power $m \neq n$ and coupling constant $\eta \neq 1$. This choice of the braiding leads to the Teledeski potential
\begin{equation}
\tilde{\mathcal{G}}(X) = M + \frac{g}{\sqrt{X}} - \frac{2 X \epsilon ^2 \left((4 m+3) (2 n+1) X^{2 n}-\eta  (2 m+1) (4 n+3) X^{2 m}\right)}{\lambda ^3 (4 m+3) (4 n+3)} \,.
\end{equation}
It is worth noting that if $m = n$ and $\eta = 1$ this reduces to $\tilde{\mathcal{G}} = M + g/\sqrt{X}$ which can be regarded to be just the $\mathcal{G}$-extension of B$\flat$M obtained earlier apart from the constant $M$. Similarly, if we consider a log-braiding ansatz inspired by the second term of Eq.~(\ref{eq:G_BFlatM}), i.e.,
\begin{equation}
G(X) = \frac{\mu}{2} \ln\left(X/X_0\right) \,,
\end{equation}
we obtain
\begin{equation}
\tilde{\mathcal{G}}(X) = \left( M - \mu \right) + \frac{g}{\sqrt{X}} - \frac{2 (2 n+1) \epsilon ^2 x^{2 n+1}}{\lambda ^3 (4 n+3)} \,.
\end{equation}
Now, invoking the linearity of the degeneracy equation (\ref{eq:BFlatM7_recast}) in $G$ and $\tilde{\mathcal{G}}$, we can add up these two cases to obtain
\begin{equation}
\label{eq:G_BFlatM7T2}
G(X) = - \frac{ \eta \epsilon^2 X^{1 + 2 m} }{ \lambda^3 } + \frac{\mu}{2} \ln\left(X/X_0\right) \,,
\end{equation}
and
\begin{equation}
\label{eq:mathcalG_BFlatM7T2}
\tilde{\mathcal{G}}(X) = \left( M - \mu \right) + \frac{g}{\sqrt{X}} - \frac{2 X \epsilon ^2 \left((4 m+3) (2 n+1) X^{2 n}-\eta  (2 m+1) (4 n+3) X^{2 m}\right)}{\lambda ^3 (4 m+3) (4 n+3)} \,.
\end{equation}
The $\mathcal{G}$-extension of B$\flat$M can be considered to be a special case $\eta = 1$, $m = n$, and $\mu = M$ of this broader form (Eqs. (\ref{eq:V_BFlatM}), (\ref{eq:G_BFlatM7T2}), and (\ref{eq:mathcalG_BFlatM7T2})).

\subsubsection*{B$\flat$TG ($G$-B$\flat$M only)}
We proceed by specifying $G$ as Eq.~(\ref{eq:G_BFlatM}). In this case, it is convenient to define the surrogate potentials $\left( \mathcal{V}, q\right)$ to $\left(V, \tilde{\mathcal{G}}\right)$ as
\begin{eqnarray}
V_X &=&\frac{\sqrt{\mathcal{V}(X)}}{\sqrt{X}} \\
\tilde{\mathcal{G}} &=& \frac{q(X)}{\sqrt{X}}\,,
\end{eqnarray}
through which the degeneracy equation simplifies to
\begin{equation}
\label{eq:degeneracy_Vq}
\lambda ^3 \sqrt{X} q_X=X \left((2 n+1) \epsilon ^2 X^{2 n}-\mathcal{V}_X\right) \,.
\end{equation}
This is obviously easier to manage since it is linear in $\mathcal{V}$ and $q$. We can close this by providing an ansatz to either $V$ or $\tilde{\mathcal{G}}$, or alternatively choose either $\mathcal{V}$ or $q$. For example, taking a power law $V$ as
\begin{equation}
V(X) = \frac{\epsilon \xi X^{1 + m} }{1 + m} \,,
\end{equation}
where $\xi$ is a constant leads to the Teledeski potential
\begin{equation}
\tilde{\mathcal{G}}(X) = \frac{g}{\sqrt{X}} + \frac{2 X \epsilon ^2 \left((4 m+3) (2 n+1) X^{2 n}-(2 m+1) (4 n+3) \xi ^2 X^{2 m}\right)}{\lambda ^3 (4 m+3) (4 n+3)} \,.
\end{equation}
Obviously, the resulting theory can be recognized as a limiting case of Eqs.~(\ref{eq:V_BFlatM}), (\ref{eq:G_BFlatM7T2}), and (\ref{eq:mathcalG_BFlatM7T2}) for $M = \mu$ due to the power law choice of $V$.

\subsubsection*{B$\flat$TG (Exponential Coupling)}

Now, we present non-power law potentials which have not yet been considered in the B$\flat$ Horndeski models and then extend them by taking the Teledeski potential into account.

We start with a well-tempered B$\flat$ Horndeski model $(\mathcal{G} = 0)$ with an exponential coupling in the kinetic term $X$. This leads us to the potentials
\begin{eqnarray}
 \label{V_exp_linear}   V(X) &=& \epsilon\,e^{2\xi X}\,,\\
 \label{G_exp_linear} G(X) &=& -\frac{4\, \epsilon^2\, \xi^2}{\lambda^3} e^{4\xi X}\, X + \frac{M}{2} \ln{ \left( X/X_0 \right)}\,,
\end{eqnarray}
where $\epsilon$ and $\xi$ are constants. It is worth noting that if $\xi \ll 1$, then this effectively reduces to the B$\flat$M potentials (\ref{eq:V_BFlatM}) and (\ref{eq:G_BFlatM}) and so the model can be regarded as a higher order effective field theory of B$\flat$M. The scalar field and Friedmann constraint on-shell are given, respectively, by
\begin{eqnarray}
    & &\phi(t) = \mp \frac{\epsilon\, \sqrt{y}\,\left[-1 + W(y)\right]}{\lambda^3 \sqrt{W(y)}}\,,\\
    & & \rho_{\Lambda} + \epsilon \, e^{W(y)/2}\, \, [-1+W(y)] = \pm \frac{\epsilon\, \sqrt{y}\, \left[-1+W(y)\right]}{\sqrt{W(y)}}\,,
\end{eqnarray}
where $W(y)$ is the Lambert $W$ function and $y:=y(t) = (-t\,\lambda^3 + 2\, \epsilon\, \xi\, c_{1})^{2} / \left( 2\, \epsilon^2\, \xi \right)$ with $c_{1}$ being an integration constant characterizing the dynamics of the scalar field. The Lambert $W(x)$ function is the solution to the transcendental equation $x = W(x) e^W(x)$ which has found many applications in physics \cite{Easther:2001fz, Easther:2001fi, Bernardo_2015, doi:10.1119/1.4983115}. We note that it is real-valued for $x > -1/e$. This makes the above parametrized on-shell solution real provided that $y(t) > -1/e$. The solution is therefore guaranteed to be real provided that $\xi >0$ and even with $\xi < 0$ as long as $y(0) > -1/e$ because $y(t)$ is a a quadratic, monotonically-increasing function of the time variable.


Next, we use the above B$\flat$ exponential Horndeski model (Eqs.~(\ref{V_exp_linear}) and (\ref{G_exp_linear})) as a basis to construct a Teledeski potentials, thereby expanding to the TG sector. Eqs.~(\ref{eq:FugueBFlatM7}-\ref{eq:FeqBFlatM7}) therefore now apply where two among the three potentials $V$, $G$, $\mathcal{G}$ must be chosen to close the system.

Taking an exponential ansatz for the braiding $G$ given by Eq.~(\ref{G_exp_linear}) and expressing the potential $V$ as a power law,
\begin{equation}
    \label{eq:V_powerlaw}
    V(X) = \alpha\, X^{\beta\, n}\,,
\end{equation}
where $\alpha$ and $\beta$ are constants, we solve the degeneracy equation (\ref{eq:BFlatM7_recast}) to obtain the Teledeski potential
\begin{eqnarray}
    \tilde{\mathcal{G}}(X) &=& -\frac{2 n^2 \alpha^2 \beta^2}{\lambda^3} \left( \frac{2\beta n-1}{4 \beta n-1} \right) X^{2\beta n-1} + \frac{\epsilon^2 \xi }{2\lambda^3} e^{4\xi X} \left(-1+8 X \xi\right) \\ \nonumber & & \quad + \frac{\epsilon^2}{8 \lambda^3} \sqrt{ \frac{\pi \xi}{X} } \rm{erfi}\left(2 \sqrt{\xi X}\right) + \frac{g}{\sqrt{X}}\,.
\end{eqnarray}
where the $\rm{erfi}(x)$ is the imaginary error function. The scalar field on-shell becomes
\begin{equation} \label{eq:exp_scalar}
    \phi(t) = c_{2} +  \frac{\alpha(1-2 \beta n)}{2^{\beta n} \lambda^3} \mathcal{P}(t)^{ 2\beta n/ \left( 2 \beta n -1 \right)} \,,
\end{equation}
 where $\mathcal{P}(t) = c_{1} \left(2 \beta n - 1\right) - 2^{\beta n - 1} \lambda^3 t / \left( \alpha \beta n \right)$
which leads to the Friedmann equation
\begin{equation} \label{eq:exp_Friedmann}
    - \frac{\alpha }{2^{\beta n}} \left(-1+2 \beta n\right) \mathcal{P}(t)^{2\beta n/ \left( 2 \beta n -1 \right)} + \rho_{\Lambda} + \lambda^3 c_{2} = 0\,.
\end{equation}
The model clearly well-tempers the vacuum energy. We note that a canonical kinetic term can also be accommodated in this model by taking the limit $\beta \rightarrow 1/n$.


We further consider a different exponentially-coupled model. This time, we take one in which the argument in the exponent is $\sqrt{x}$ and build on the resulting B$\flat$ Horndeski model $(\mathcal{G} = 0)$. We are then lead to the Horndeski potentials
\begin{eqnarray}
    V(X) &=& \epsilon\, e^{\xi \sqrt{2 X}}\,,\\
 \label{G_exp_root} G(X) &=& -\frac{\epsilon^2\, \xi^2\, e^{2 \xi \sqrt{2 X}}}{2 \lambda^3} + \frac{M}{2} \ln{X}\,,
\end{eqnarray}
where $\epsilon$ and $\xi$ are constants. In contrast with the previous non-power law potentials, these clearly do not reduce to the Galileon in any limit of its parameter space. The scalar field and Friedmann constraint on-shell are given by
\begin{eqnarray}
    \label{eq:phi_on_shell_exproot} & &\phi(t) =  -\frac{\epsilon}{\lambda^3}\left[ z - z\, \ln{z} - \xi\, c_1\right]\,,\\
    \label{eq:Hc_on_shell_exproot} & &\rho_{\Lambda} = \epsilon\, \xi\, c_1\,,
\end{eqnarray}
where $z:=z(t) = \left( \epsilon\, \xi^2\, c_1 - \lambda^3\, t \right) / \left( \epsilon \xi \right)$ and $c_{1}$ is the integration constant. As with the previous example, we now use this as a basis to construct an exponential Teledeski model. 


Extending the above B$\flat$ Horndeski model, we take in its braiding potential (\ref{G_exp_root}) and a Horndeski $V$ potential (\ref{eq:V_powerlaw}). Substituting these into the degeneracy equation (\ref{eq:BFlatM7_recast}), we obtain the Teledeski potential
\begin{equation}
    \tilde{\mathcal{G}}(X) = \frac{2 n^2 \alpha^2 \beta^2 }{\lambda^3} \left( \frac{2n\beta - 1}{4n\beta - 1} \right) X^{ \left( 4n\beta - 3\right)/2} - \frac{ \epsilon^2 \xi }{8\lambda^3 } e^{2\sqrt{2X} \xi} \left( \sqrt{\frac{2}{X}} - 4 \xi \right) + \frac{g}{\sqrt{X}}\,.
\end{equation}
The on-shell scalar and Friedmann constraint are given by Eqs.~(\ref{eq:phi_on_shell_exproot}) and (\ref{eq:Hc_on_shell_exproot}), the same ones as the corresponding B$\flat$ Horndeski model since the on-shell behavior is explicitly dependent on $V$ due to well-tempering. We lastly note that this exponential coupling model may accommodate a canonical kinetic term in the limit $\beta \rightarrow 1/n$.

\subsubsection*{B$\flat$TG (Arctan Coupling)}

As a last example, and an illustration to how complicated the well-tempered models can become as one deviates away from the aesthetic power law (Galileon) ansatzes, we consider an $\arctan$ coupling in the kinetic term.

We fix the braiding to be Eq.~(\ref{eq:G_BFlatM}). The degeneracy equation then reduces to Eq.~(\ref{eq:degeneracy_Vq}) but this time we take in the Horndeski potential given by
\begin{equation}
V(X) = \xi \arctan\left( X / \chi \right)\,,
\end{equation}
where $\xi$ and $\chi$ are constants. Substituting this into the degeneracy equation (\ref{eq:degeneracy_Vq}), we obtain the Teledeski potential $\tilde{\mathcal{G}}(X) = \left( g/\sqrt{X} \right) + \epsilon^2 \mathcal{Y}(X) / \left( \lambda^3 \sqrt{X} \right)$ where $\mathcal{Y}(X)$ is given by
\begin{eqnarray}
\mathcal{Y}(X) &=& \phantom{+} \frac{2 (2 n+1) }{4 n+3} X^{(4n + 3)/2} \, _2F_1\left(3,n+\frac{3}{4};n+\frac{7}{4};-\frac{X^2}{\chi ^2}\right) \nonumber\\
& & + \frac{6 (2 n+1) }{\chi ^2 (4 n+7)} X^{(7 + 4n)/2} \, _2F_1\left(3,n+\frac{7}{4};n+\frac{11}{4};-\frac{X^2}{\chi ^2}\right) \nonumber\\
& & + \frac{6 (2 n+1) }{\chi ^4 (4 n+11)} X^{(11 + 4n)/2} \, _2F_1\left(3,n+\frac{11}{4};n+\frac{15}{4};-\frac{X^2}{\chi ^2}\right) \nonumber\\
& & + \frac{2 (2 n+1)  }{\chi ^6 (4 n+15)} X^{(15 + 4n)/2} \, _2F_1\left(3,n+\frac{15}{4};n+\frac{19}{4};-\frac{X^2}{\chi ^2}\right) \nonumber\\
& & + \frac{\xi^2}{32} \bigg[ \ 8 X^{3/2} \frac{ \left(X^2-3 \chi ^2\right)}{\left(X^2 + \chi ^2\right)^2} +  \sqrt{ \frac{2}{\chi } } \ln \left( \frac{ \chi -\sqrt{2} \sqrt{\chi } \sqrt{X}+X }{\chi +\sqrt{2} \sqrt{\chi } \sqrt{X}+X} \right) \nonumber\\
& & \phantom{gggggg} + 2 \sqrt{ \frac{2}{\chi } } \left[\arctan\left( 1 + \sqrt{\frac{2X}{\chi }}\right)-\arctan\left(1-\sqrt{\frac{2X}{\chi }}\right)\right] \ \bigg]\,,
\end{eqnarray}
where $_2F_1(a, b; c; z)$ is the hypergeometric function. The Hamiltonian constraint becomes
\begin{equation}
\rho_\Lambda +\lambda ^3 \phi (t)+\frac{2 \xi  \chi  \epsilon X}{ X^2+ \chi ^2} - \xi  \epsilon  \arctan\left(\frac{X}{\chi }\right) = 0\,,
\end{equation}
and provides assurance that a well-tempered Minkowski vacuum is accommodated. We must also mention that an exact analytical solution to the scalar field on-shell is available in this model; however, it is not presentable. Nonetheless, this confirms that the $\arctan$ coupling model derived above well-tempers.

\section{Functionals of the Well-Tempered Teleparallel Galileon}
\label{sec:UW_galileon}

We present the explicit functional expressions $Q$, $S$, $U$, and $W$ which define the dynamical system (Eqs.~(\ref{eq:chip_galileon}) and (\ref{eq:yp_galileon})) of the well-tempered teleparallel Galileon (Sec.~\ref{subsec:phase_portraits})
\begin{eqnarray}
    Q[\chi, y] &=& \bigg[  48 \chi ^{3/2} \left[\alpha^2 \chi +12 \sqrt{2} \sigma  y^3 \left(2 \sqrt{\chi } \left(4 \chi  \left(\alpha ^2+\gamma \right)-3 M\right)-3 g\right)-6 \sqrt{2} \zeta  \chi ^{3/2} y\right] \nonumber\\
    & & \times \bigg(\sqrt{2} \alpha  \sqrt{\chi }+2 y \bigg(-3 \alpha ^2 \chi +M+9 \sqrt{2} \zeta  \chi ^{3/2} y \nonumber\\
    & & \phantom{ggggggggggggggggi} +9 \sqrt{2} \sigma  y^3 \left(3 g+2 \sqrt{\chi } \left(3 M-4 \chi  \left(\alpha ^2+\gamma \right)\right)\right) \bigg) \bigg) \nonumber\\
    & & - \frac{8 \sqrt{2} \chi}{3 M y^2-1} \bigg[1 + 3 y \bigg[\sqrt{2} \alpha  \sqrt{\chi }+y \bigg(-6 \alpha ^2 \chi+M+18 \sqrt{2} \zeta  \chi ^{3/2} y \nonumber\\
    & & \phantom{gggggggggggggggggg} +18 \sqrt{2} \sigma  y^3 \left(3 g+2 \sqrt{\chi } \left(3 M-4 \chi  \left(\alpha ^2+\gamma \right)\right)\right)\bigg)\bigg]\bigg] \nonumber\\
    & & \times \bigg(36 y^4 \left(6 g M \sigma  \sqrt{\chi }+3 M^2 \sigma  \chi +2 M \sigma  \chi ^2 \left(\alpha ^2+\gamma \right)\right) \nonumber\\
    & & \phantom{gggg} -36 y^2 \left(6 g \sigma  \sqrt{\chi }+\chi ^2 \left(\zeta  M-4 \sigma  \left(\alpha ^2+\gamma \right)\right)+6 M \sigma  \chi \right) \nonumber\\
    & & \phantom{gggg} +1 -l^2 M-\chi  (3 \zeta  \chi +\alpha  M)+6 \sqrt{2} \alpha ^2 M \chi ^{3/2} y\bigg) \bigg] \,,
\end{eqnarray}

\begin{eqnarray}
    S[\chi, y] &=& \bigg[  648 \sigma ^2 y^7 \bigg(224 \chi ^3 \left(\alpha ^2+\gamma \right)^2+18 g^2-132 g \chi ^{3/2} \left(\alpha ^2+\gamma \right) \nonumber\\
    & & \phantom{gggggggg} +81 g M \sqrt{\chi }+90 M^2 \chi -288 M \chi ^2 \left(\alpha ^2+\gamma \right)\bigg) \nonumber\\
    & & +18 \sqrt{2} \sigma  y^4 \bigg(240 \alpha ^2 \chi ^{5/2} \left(\alpha ^2+\gamma \right)-63 \alpha ^2 g \chi  \nonumber\\
    & & \phantom{ggggggggggg} +6 g M +18 M^2 \sqrt{\chi }-8 M \chi ^{3/2} \left(23 \alpha ^2+5 \gamma \right)\bigg) \nonumber\\
    & & +972 y^5 \left(-32 \zeta  \sigma  \chi ^3 \left(\alpha ^2+\gamma \right) +9 \zeta  g \sigma  \chi ^{3/2}+20 \zeta  M \sigma  \chi ^2 \right) \nonumber\\
    & & +18 y^3 \left(-64 \alpha  \sigma  \chi ^2 \left(\alpha ^2+\gamma \right)+90 \zeta ^2 \chi ^3+15 \alpha  g \sigma  \sqrt{\chi }+36 \alpha  M \sigma  \chi \right) \nonumber\\
    & & +12 \sqrt{2} y^2 \left(3 g \sigma -2 \sqrt{\chi } \left(\chi  \left(4 \sigma  \left(\alpha ^2+\gamma \right)+18 \alpha ^2 \zeta  \chi -3 \zeta  M\right)-3 M \sigma \right)\right) \nonumber\\
    & & +\sqrt{2} \sqrt{\chi } \left(6 \chi  \left(\zeta -2 \alpha ^3\right)+\alpha  M\right)+3 \alpha  \chi  y \left(18 \chi  \left(\alpha ^3+2 \zeta \right)-5 \alpha  M\right) \bigg] \,,
\end{eqnarray}

\newpage

\begin{eqnarray}
    U[\chi, y] &=& 4 \sqrt{\chi } \bigg[  \frac{\sqrt{2}}{3 M y^2-1} \bigg[-\sqrt{2} \alpha -6 y \bigg(9 \sqrt{2} y \left(\zeta  \chi +2 \sigma  y^2 \left(M-4 \chi  \left(\alpha ^2+\gamma \right)\right)\right)-2 \alpha ^2 \sqrt{\chi }\bigg)\bigg] \nonumber\\
    & & \times \bigg(36 y^4 \left(6 g M \sigma  \sqrt{\chi }+3 M^2 \sigma  \chi +2 M \sigma  \chi ^2 \left(\alpha ^2+\gamma \right)\right) \nonumber\\
    & & \phantom{gggg}-36 y^2 \left(6 g \sigma  \sqrt{\chi }+\chi ^2 \left(\zeta  M-4 \sigma  \left(\alpha ^2+\gamma \right)\right)+6 M \sigma  \chi \right) \nonumber\\
    & & \phantom{gggg}+ 1 -l^2 M-\chi  (3 \zeta  \chi +\alpha  M)+6 \sqrt{2} \alpha ^2 M \chi ^{3/2} y \bigg)\nonumber \\
    & & +12 \left(\alpha ^2 \chi +12 \sqrt{2} \sigma  y^3 \left(2 \sqrt{\chi } \left(4 \chi  \left(\alpha ^2+\gamma \right)-3 M\right)-3 g\right)-6 \sqrt{2} \zeta  \chi ^{3/2} y\right)^2 \bigg]\,,\label{eq:U_galileon}
\end{eqnarray}
and
\begin{eqnarray}
    W[\chi, y] &=& -36 \sqrt{2} \alpha ^2 \chi  y^3 \left(24 g \sigma +\sqrt{\chi } \left(8 \sigma  \chi  \left(\alpha ^2+\gamma \right)+3 M (7 \zeta  \chi +4 \sigma )\right)\right) \nonumber\\
    & & +\alpha  \left(-1 +l^2 M+\chi  \left(-6 \alpha ^3 \chi +3 \zeta  \chi +\alpha  M\right)\right) \nonumber\\
    & & -6 \sqrt{2} \alpha ^2 \sqrt{\chi } y \left(-1 +l^2 M+\chi  (2 \alpha  M-9 \zeta  \chi )\right) \nonumber\\
    & & -1296 \sqrt{2} \alpha ^2 M \sigma  \chi ^{3/2} y^5 \left(2 M-5 \chi  \left(\alpha ^2+\gamma \right)\right) \nonumber\\
    & & + 1296 M \sigma ^2 y^8 \bigg(36 g^2+6 g \sqrt{\chi } \left(21 M-20 \chi  \left(\alpha ^2+\gamma \right)\right) \nonumber\\
    & & \phantom{gggggggggggggi} +\chi  \left(280 \chi ^2 \left(\alpha ^2+\gamma \right)^2+135 M^2-354 M \chi  \left(\alpha ^2+\gamma \right)\right)\bigg)\nonumber\\
    & & + 216 \sigma  y^6 \bigg(-72 g^2 \sigma -6 g \sqrt{\chi } \left(8 \sigma  \chi  \left(\alpha ^2+\gamma \right)+M (30 \sigma -3 \zeta  \chi )\right) \nonumber\\
    & & \phantom{gggggggggg}+\chi \bigg(-224 \sigma  \chi ^2 \left(\alpha ^2+\gamma \right)^2+45 M^2 (3 \zeta  \chi -4 \sigma )\nonumber\\
    & & \phantom{ggggggggggggggg}-6 M \chi  \left(\alpha ^2+\gamma \right) (47 \zeta  \chi -44 \sigma ) \bigg)\bigg) \nonumber\\
    & & +36 y^4 \bigg(12 \sigma  \chi  \left(\left(\alpha ^2+\gamma \right) \left(11 \zeta  \chi ^2+1 \right)+15 \zeta  g \sqrt{\chi }\right) \nonumber\\
    & & \phantom{ggggggg} +3 l^2 M \sigma  \left(M-4 \chi  \left(\alpha ^2+\gamma \right)\right) \nonumber\\
    & & \phantom{ggggggg} +M \left(\sigma  \chi ^2 \left(45 \zeta -14 \alpha  \left(\alpha ^2+\gamma \right)\right)+90 \zeta ^2 \chi ^3-6 \alpha  g \sigma  \sqrt{\chi }- 3 \sigma \right) \bigg) \nonumber\\
    & & +18 y^2 \bigg(12 \alpha  g \sigma  \sqrt{\chi }+\chi  \bigg(-15 \zeta ^2 \chi ^2+3 \left(- \zeta  +\zeta  l^2 M+4 \alpha  M \sigma \right) \nonumber\\
    & & \phantom{ggggggggggggggggggggg} +\alpha  \chi  \left(5 M \left(\alpha ^3+\zeta \right)-8 \sigma  \left(\alpha ^2+\gamma \right)\right)\bigg)\bigg) \,.\label{eq:W_galileon}
\end{eqnarray}

\section*{References}

\providecommand{\href}[2]{#2}\begingroup\raggedright\endgroup

\end{document}